\documentclass[a4paper,11pt]{article}
\pdfoutput=1 

\usepackage{jcappub} 

\usepackage{multirow}
\usepackage{xcolor}
\usepackage{booktabs}
\usepackage{bm}

\newcommand{\DM}{{\rm DM}}
\newcommand{\pccc}{\,pc\,cm$^{-3}$\,}
\newcommand{\kms}{km s$^{-1}$ Mpc$^{-1}$}

\title{Probing the interaction between dark energy and dark matter with future fast radio burst observations}

\author[a]{Ze-Wei Zhao,}
\author[a]{Ling-Feng Wang,}
\author[a]{Ji-Guo Zhang,}
\author[a]{Jing-Fei Zhang}
\author[a,b,c,1]{and Xin Zhang\note{Corresponding author.}}
\affiliation[a]{Key Laboratory of Cosmology and Astrophysics (Liaoning) \& Department of Physics, College of Sciences, Northeastern
University, Shenyang 110819, China}
\affiliation[b]{Key Laboratory of Data Analytics and Optimization for Smart Industry
(Ministry of Education), Northeastern University, Shenyang 110819, China}
\affiliation[c]{National Frontiers Science Center for Industrial Intelligence and Systems Optimization,
Northeastern University, Shenyang 110819, China}

\emailAdd{
zhaozw@stumail.neu.edu.cn, lingfengwang@stumail.neu.edu.cn, zhangjiguo@stumail.neu.edu.cn, jfzhang@mail.neu.edu.cn, zhangxin@mail.neu.edu.cn}

\abstract{Interacting dark energy (IDE) scenario assumes that there exists a direct interaction between dark energy and cold dark matter, but this interaction is hard to be tightly constrained by the current data. Fast radio bursts (FRBs) will be seen in large numbers by future radio telescopes, and thus they have potential to become a promising low-redshift cosmological probe. In this work, we investigate the capability of future FRBs of constraining the dimensionless coupling parameter $\beta$ in four phenomenological IDE models. If we fix the FRB properties, about $10^5$ FRB data can give constraints on $\beta$ tighter than the current cosmic microwave background data in the IDE models with the interaction proportional to the energy density of dark energy. In all the IDE models, about $10^6$ FRB data can achieve the absolute errors of $\beta$ to less than $0.10$, providing a way to precisely measure $\beta$ by only one cosmological probe. Jointly constraining the FRB properties and cosmological parameters would increase the constraint errors of $\beta$ by a factor of about 0.5--2.
}

\begin{document}

\maketitle

\section{Introduction} \label{sec:intro}
Fast radio bursts (FRBs) are a class of astronomical radio pulses with high energy. An FRB could interact with free electrons in the plasma and generate dispersion, resulting in its lower frequency signal being delayed. The amount of dispersion is determined by the free electron number density along the line of sight and can be quantified by dispersion measure (DM). Observed FRBs all have high DMs, exceeding the contribution from the Milky Way. This fact indicates that most FRBs are likely of extragalactic origin, except that one FRB event has been detected within the Milky Way \cite{CHIMEFRB:2020abu,Bochenek:2020zxn}. The value of DM contains the baryonic and distance information along the FRB signal's path, so we can measure cosmological parameters by the Macquart relation, which describes the relationship between $\rm DM$ and redshift $z$ \cite{Macquart:2020lln}. A lot of works have proposed that the FRB data can constrain cosmological parameters \cite{Deng:2013aga,Gao:2014iva,Zhou:2014yta,Yang:2016zbm,Li:2017mek,Walters:2017afr,Jaroszynski:2018vgh,Liu:2019jka,Liu:2019ddm,Zhang:2020btn,Qiang:2021bwb,Dai:2021czy,Zhao:2021jeb,Zhu:2022mzv,Wu:2022dgy} and related parameters \cite{Li:2019klc,Wei:2019uhh,Wu:2020jmx,Lee:2021ppm,Gao:2022ifq,Reischke:2023gjv}.
Recent reviews and other cosmological applications can refer to refs.~\cite{Bhandari:2021thi,Xiao:2021omr,Petroff:2021wug,Caleb:2021xqe} and references therein.

So far, hundreds of FRB events have been detected, with more than 600 events reported by the Canadian Hydrogen Intensity Mapping Experiment (CHIME)/FRB project \cite{CHIMEFRB:2021srp}, but only a few events are localized to known host galaxies and with certain redshifts. As a result, current researches about using FRBs to measure cosmological parameters mainly focus on constraining the cosmic baryon density and the Hubble constant $H_0$ in the base $\Lambda$ cold dark matter ($\Lambda$CDM) model. Macquart et al. \cite{Macquart:2020lln} used five localized FRBs observed by the Australian Square Kilometre Array Pathfinder (ASKAP) to derive the constraint on the cosmic baryon density with a precision of around 40\%, which is consistent with the result from the big bang nucleosynthesis and the cosmic microwave background (CMB) measurements. For measuring $H_0$ using FRBs, several independent groups \cite{Hagstotz:2021jzu,Wu:2021jyk,James:2022dcx,Liu:2022bmn,Zhao:2022yiv} used different methods and obtained the constraints which are consistent at the 1$\sigma$ confidence level.

However, for the dark energy parameters in the extended $\Lambda$CDM models, current few localized FRB events cannot provide tight constraints. There is no doubt that much more localized FRBs will be detected in the future. The construction of CHIME Outrigger telescopes\footnote{https://chime-experiment.ca/en} and the Commensal Real-time ASKAP Fast Transients Coherent upgrade project\footnote{https://www.atnf.csiro.au/projects/askap/index.html} are in progress and the Square Kilometre Array (SKA) project is also in construction. It is predicted that about $10^3$--$10^6$ FRB data could be accumulated in several years by these telescopes \cite{Bhandari:2021thi}. Previous works forecast that for the $w$CDM and Chevallier--Polarski--Linder (CPL) models, about $10^4$ localized FRBs can constrain the dark energy parameters with the same precision as the CMB data \cite{Zhao:2020ole}, while for the holographic dark energy and Ricci dark energy models, more than $10^4$ localized FRBs are needed \cite{Qiu:2021cww}.

These works show that FRBs are a promising probe to constrain dark energy parameters, but they mainly focus on the properties of dark energy and only study how the FRB data can constrain the equation of state (EoS) of dark energy.
In fact, dark energy may have a direct interaction with cold dark matter, which can be described by the interacting dark energy (IDE) models
(see,
e.g.,  ref. \cite{Wang:2016lxa} for a recent review). The IDE models could help resolve some theoretical and observational problems, such as the cosmic coincidence problem \cite{Comelli:2003cv,Cai:2004dk,Zhang:2005rg,He:2008tn,He:2009pd} and the $H_0$ tension \cite{DiValentino:2017iww,Yang:2018euj,Yang:2018uae,Pan:2019gop,DiValentino:2019ffd,DiValentino:2019jae,Vagnozzi:2019ezj,Gao:2021xnk}. Because the microscopic nature of dark energy and dark matter is still unclear, the energy transfer rate in the IDE models can be considered in a purely phenomenological way \cite{Zhang:2007uh,Zhang:2009qa,Li:2009zs,Li:2010ak,Li:2011ga,Zhang:2012uu,Zhang:2013lea,Li:2013bya,Li:2014eha,Li:2014cee,Geng:2015ara,Li:2015vla}, i.e., proportional to the energy density of dark energy, dark matter, or some mixture of them, and the proportionality coefficient is commonly considered as $\beta H$ or $\beta H_0$. Here, $\beta$ is a dimensionless coupling parameter \cite{Amendola:1999qq,Billyard:2000bh} and $H$ is the Hubble parameter.

One can use cosmological data to constrain the parameter $\beta$ in the IDE models to indirectly detect the interaction between dark sectors \cite{Guo:2007zk,Xia:2009zzb,He:2010im,Fu:2011ab,Murgia:2016ccp,Costa:2016tpb,Feng:2016djj,Xia:2016vnp,Guo:2018gyo,Li:2018ydj,Feng:2019jqa,Cheng:2019bkh,Aljaf:2020eqh,Li:2020gtk,Zhang:2021yof,Lucca:2021eqy,Nunes:2022bhn}. However, current observational data cannot precisely constrain $\beta$  \cite{Yang:2021hxg,Guo:2021rrz,Wang:2021kxc,Jin:2022tdf}.
This is because the interaction between dark sectors mainly affects the evolution of the late-time Universe, but the current late-Universe cosmological probes are not precise enough. Measuring this interaction could deepen our understanding of dark energy and dark matter. Therefore, it is necessary to find and evaluate new late-Universe cosmological probes by using mock data. Now that FRBs have displayed fine capability on constraining dark energy parameters in some dark energy models, they may also help constrain the coupling parameter $\beta$ in the IDE models. To study how many localized FRBs could provide precise measurement on $\beta$, in this paper, we simulate future FRB data to constrain four typical IDE models.

Our paper is organized as follows. In Section \ref{sec:Method}, we briefly describe the IDE models and the methods of simulating the FRB data. The constraints and relevant discussion are given in Section \ref{sec:Result}. Conclusion is given in Section \ref{sec:con}. Throughout this paper, we adopt the units in which the speed of light equals 1.

\section{Methods and data}\label{sec:Method}

\subsection{Brief description of the IDE models}
In a spatially flat Friedmann--Roberston--Walker universe, the Friedmann equation can be written as
\begin{equation}\label{2.1}
3M^2_{\rm{pl}} H^2=\rho_{\rm{de}}+\rho_{\rm c}+\rho_{\rm b}+\rho_{\rm r},
\end{equation}
where $3M^2_{\rm{pl}} H^2$ is the critical density of the Universe and $\rho_{\rm{de}}$, $\rho_{\rm c}$, $\rho_{\rm b}$, and $\rho_{\rm r}$ represent the energy densities of dark energy, cold dark matter, baryon, and radiation, respectively.

If we assume that there exists a direct interaction between dark energy and cold dark matter, we can obtain the energy conservation equations,
\begin{align}\label{conservation1}
\dot{\rho}_{\rm de} +3H(1+w)\rho_{\rm de}= Q,\\
\dot{\rho}_{\rm c} +3 H \rho_{\rm c}=-Q,
\end{align}
where the dot denotes the derivative with respect to the cosmic time $t$, $w$ is the EoS of dark energy, and $Q$ is the energy transfer rate.

We consider the forms of $Q$ in a purely phenomenological way and choose four typical forms, i.e., $Q=\beta H\rho_{\rm c}$ (IDE1), $Q=\beta H_0\rho_{\rm c}$ (IDE2), $Q=\beta H\rho_{\rm de}$ (IDE3), and $Q=\beta H_0\rho_{\rm de}$ (IDE4).
Here $\beta > 0$ indicates cold dark matter decaying into dark energy, and $\beta < 0$ indicates dark energy decaying into cold dark matter. $\beta = 0$ indicates no interaction between dark energy and cold dark matter.
Because we focus on how the FRB data could constrain $\beta$, we do not wish to introduce more extra parameters, and thus for the EoS of dark energy, we only consider $w=-1$ (corresponding to the vacuum energy).

In the IDE models, since vacuum energy is not a true background, we need to consider the perturbations of vacuum energy.
In the standard linear perturbation theory, dark energy is considered as a nonadiabatic fluid with negative pressure. Once dark energy interacts with cold dark matter, the interaction would affect the nonadiabatic pressure perturbation of dark energy and further make the nonadiabatic curvature perturbation occasionally diverge on the large scales, known as the large-scale instability problem \cite{Majerotto:2009zz,Clemson:2011an,He:2008si}. To avoid this problem, one needs to use a proper approach to treat the perturbations of dark energy. In 2014, Li, Zhang, and Zhang~\cite{Li:2014eha,Li:2014cee} extended the parametrized post-Friedmann (PPF) approach \citep{Fang:2008sn,Hu:2008zd} to the IDE models, referred to as the ePPF approach. This approach can successfully avoid the large-scale instability problem in the IDE models. In this work, we employ the ePPF method to treat the cosmological perturbations; see, e.g.,  refs. \cite{Zhang:2017ize,Feng:2018yew} for more applications of the ePPF method.

\subsection{Mock FRB data}\label{21}
The DM of an FRB can be measured by the time delay of the signal between the highest frequency and the lowest frequency. The value of DM equals the integral of the electron number density $n_{\mathrm{e}}$ weighted by $(1+z)^{-1}$, along the path to this FRB,
\begin{eqnarray}
\mathrm{D M}=\int \frac{n_{\mathrm{e}}(l)}{1+z} \mathrm{d} l.
\end{eqnarray}

The observed DM of an FRB can be modelled by the sum of the contributions from the Milky Way's interstellar
medium and halo, the intergalactic medium (IGM), and the FRB host galaxy,
\begin{eqnarray} \label{DMcom}
\DM  =  \DM_{\rm ISM}+ \DM_{\rm halo} + \DM_{\rm IGM} + \DM_{\rm host}.
\end{eqnarray}
The average value of the cosmic contribution $\rm{DM}_{\rm{IGM}}$ is expressed as
\begin{align}\label{eq3}
\langle\mathrm{DM}_{\mathrm{IGM}}\rangle=\frac{3H_0\Omega_{\rm b}f_{\mathrm{IGM}}}{8\pi G m_{\mathrm{p}}}\int_0^z\frac{\chi(z')(1+z')dz'}{E(z')},
\end{align}
with
\begin{align}
\chi(z)=Y_{\rm H}\chi_\mathrm{{e,H}}(z)+\frac{1}{2}Y_{\rm He}\chi_\mathrm{{e,He}}(z),
\end{align}
where $f_{\mathrm{IGM}}\simeq0.83$  is the fraction of baryon mass in the IGM \cite{Shull:2011aa},
$\Omega_\mathrm{b}$ is the present-day baryon fractional density, $G$ is Newton's constant, $m_{\mathrm{p}}$ is the proton mass, $Y_{\rm H}=3/4$ is the hydrogen mass fraction, and $Y_{\rm He}=1/4$ is the helium mass fraction. The terms $\chi_\mathrm{{e,H}}$ and $\chi_\mathrm{{e,He}}$ are the ionization fractions for H and He, respectively. We take $\chi_\mathrm{{e,H}}=\chi_\mathrm{{e,He}}=1$ \cite{Fan:2006dp}, since both H and He can be regarded as fully ionized within $z<3$.

The distribution of baryon matter in the IGM is inhomogeneous, so for an individual FRB, $\rm{DM}_{\rm{IGM}}$ is scattered from the mean value. Here we use the power-law form of the uncertainty of $\rm{DM}_{\rm{IGM}}$ fitted from the results in ref. \cite{McQuinn:2013tmc},
\begin{equation}
\sigma_{\rm{IGM}}(z)=173.8~z^{0.4}~\rm{pc}~\rm{cm}^{-3},
\end{equation}
where the top hat model for halos’ gas profile of the ionized baryons is assumed.

The value of $\DM_{\rm ISM}$ is estimated by the NE2001 model\footnote{Ben Bar-Or, J.~Prochaska, available at https://readthedocs.org/projects/ne2001/},  $\DM_{\rm halo}$ is set to be DM$_{\rm halo}=50$\,pc\,cm$^{-3}$, and their uncertainties can be absorbed in the uncertainty of $\DM_{\rm host}$ \cite{Macquart:2020lln,James:2022dcx}. From eq.~(\ref{DMcom}), we can see that if we further ignore the uncertainty of  $\rm{DM}_{\rm{obs}}$, which is enough small relative to the other errors, the total uncertainty is determined by
\begin{align}\label{eq6}
\sigma_{\rm{DM}}=\left[\sigma_{\rm IGM}^{2}
+\left(\frac{\sigma_{\rm host}}{1+z}\right)^{2} \right]^{1/2}.
\end{align}
Following ref. \cite{Li:2019klc}, we assume that the uncertainty of ${\rm DM_{host}}$ would achieve $\sigma_{\rm{host}} = 30 ~{\rm {pc~cm^{-3}}}$. If the values of $\DM_{\rm ISM}$, $\DM_{\rm halo}$, and $\DM_{\rm host}$ are known {and we adopt the assumption that all the FRB data are independent}, then the $\chi^2$ function is simply written as
\begin{align}
\chi_{\rm cos}^2=\sum_{i=1}^{N}\left[\frac{\DM_i-\langle\DM_{\rm IGM}\rangle(z_i;\bm{\theta})}{\sigma_{\rm DM}(z_i)}\right]^{2},
\end{align}
where $z_i$ and $\DM_i$ represent the redshift and DM of the $i$-th mock FRB data, respectively, $N$ is the total event number, and $\bm{\theta}$ denotes cosmological parameters, including the six base $\Lambda$CDM cosmological parameters and an extra IDE parameter $\beta$. The Hubble constant $H_0$ and the present-day matter density $\Omega_{\rm m}$ are derived parameters. Then we treat $\bm{\theta}$ as free parameters and use the Markov Chain Monte Carlo (MCMC) method to constrain them. {It should be noted that actually the DM values of FRB data are correlated due to long wavelength modes of the electron distribution of the large-scale structure \cite{Reischke:2023blu}. The effect of the correlation on cosmological constraints depends on the sky area and the number of events of the observed FRB data. A more accurate calculation of the FRB likelihood should include the survey area of different telescopes and the covariance between different FRB data. We leave this study to future work.}

The method above focuses on only constraining cosmological parameters and depends on some assumptions of $\DM_{\rm host}$, i.e., the mean value of $\DM_{\rm host}$ is fixed to the best-fit constraint values or the values from cosmological simulations \cite{Zhang:2020mgq}. Jointly constraining the FRB properties and cosmological
parameters is more realistic by considering their correlations.

For simplicity, we assume that the distribution of $\DM_{\rm host}$ can be approximated as a Gaussian function with the fiducial mean value $\mu_{\rm fid}=100$ \pccc and standard deviation 20 \pccc. According to this distribution, we randomly select a $\DM_{\rm host}$ value for each FRB event and add it to the DM value, with the redshift correction $1/(1+z)$. If $\mu$ has an intrinsic redshift evolution, we assume a power-law evolution $\mu(z)=\mu (1+z)^\gamma$ and set $\gamma_{\rm fid}=0.7$. We model the $\chi^2$ function as
\begin{align}
\chi_{\rm joint}^2=\sum_{i=1}^{N}\left[\frac{\DM_i-\frac{\mu (1+z)^\gamma}{1+z}-\langle\DM_{\rm IGM}\rangle(z_i;\bm{\theta})}{\sigma_{\rm DM}(z_i)}\right]^{2}.
\end{align}
We first assume the $\DM_{\rm host}$ distribution does not evolve with redshift (i.e. $\gamma=0$) and treat $\mu$ and $\bm{\theta}$ as free parameters. Then we consider the redshift evolution of $\DM_{\rm host}$ by treating all the DM parameters (i.e. $\mu$ and $\gamma$) and cosmological parameters $\bm{\theta}$ as free parameters and use MCMC to jointly constrain them.

In the era of SKA, it is expected that $10^5$-$10^6$ FRBs could be detected per year \cite{Hashimoto:2020dud}. Assuming that the observation duration of SKA is 10-year and 1/10 of FRBs can be localized to the host galaxy with redshift, we consider a conservative expectation with the mock FRB event number $N_{\rm FRB}=10^5$ and an optimistic expectation with $N_{\rm FRB}=10^6$ (a detailed discussion about the FRB event number can refer to ref. \cite{ZJG}).

\begin{figure}[htbp]
{\includegraphics[width=0.8\linewidth]{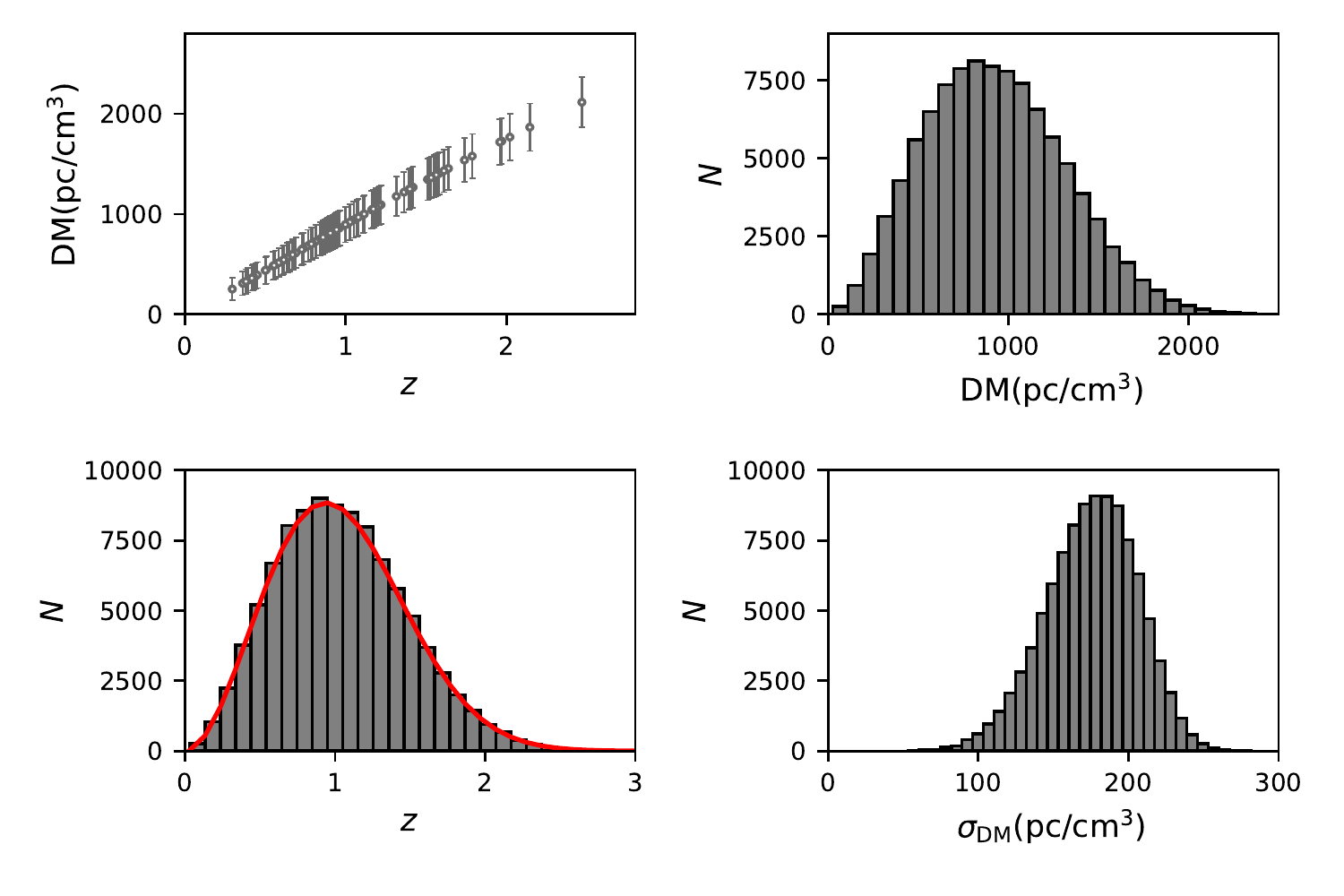}
}
\centering
\caption{The statistical properties of mock FRB data with $N_{\rm{FRB}} = 10^{5}$ in the fiducial IDE1 cosmology. Top left panel: the representative FRB events with $1\sigma$ errors. We show one data point for every 1000 events. Top right panel: The histogram of FRB event number versus ${\rm DM}$. Bottom left panel: The redshift evolution of FRB event number. The assumed probability distribution function of redshift is also shown by solid red line. Bottom right panel: The histogram of FRB event number versus $\sigma_{\rm DM}$.}
 \label{DM}
\end{figure}

Except for a Galactic magnetar, the progenitors of FRBs have not been generally identified. Therefore, the accurate redshift distribution of FRBs is unknown. Based on the first CHIME/FRB catalog, the common assumption that FRBs' population tracks the star formation history (SFH) of the Universe has been ruled out \cite{Zhang:2021kdu}. Qiang et al. \cite{Qiang:2021ljr} also confirmed this result, and proposed some empirical distribution models consistent with the observational data. We use the simple power-law model of redshift distribution to simulate the FRB data. The event rate of mock FRB data is
\begin{align}
N_{\rm{SFH}}(z)=(1+z)^\gamma\mathcal{N}_{\rm{SFH}}\frac{\dot{\rho}_{*}{d^2_{\rm C}}(z)}{H(z)(1+z)}e^{-{d^2_{\rm{L}}}(z)/[2{d^2_{\rm{L}}}(z_{\rm cut})]},
\end{align}
where $(1+z)^\gamma$ represents the delay with respect to SFH with $\gamma=-1.1$ \cite{Qiang:2021ljr}, $\mathcal{N}_{\rm{SFH}}$ is a normalization factor, $d_{\rm C}$ is the comoving distance, $d_{\rm L}$ is the luminosity distance, and $z_{\rm cut}=1$ is a Gaussian cutoff above which the number of detected FRBs would decrease due to the detection threshold. The form of SFH is \cite{Madau:2016jbv,Qiang:2021ljr}
\begin{align}
\dot{\rho}_{*}(z)=\frac{(1+z)^{2.6}}{1+((1+z)/3.2)^{6.2}}.
\end{align}
The statistical properties of the mock FRB data with $N_{\rm{FRB}} = 10^{5}$ in the fiducial IDE1 cosmology are shown in figure \ref{DM}.

\subsection{Cosmological data}

For the current cosmological data as comparison, we use the \emph{Planck} 2018 CMB ``distance priors"  \cite{Chen:2018dbv}, and the baryon acoustic oscillation (BAO) measurements from 6dFGS at $z_{\rm eff} = 0.106$ \cite{Beutler:2011hx}, SDSS-MGS at $z_{\rm eff} = 0.15$ \cite{Ross:2014qpa}, and BOSS-DR12 at $z_{\rm eff} = 0.38$, 0.51, and 0.61 \cite{Alam:2016hwk}. For the type Ia supernova (SN) data, we use the sample from the Pantheon compilation with 1048 data \cite{Pan-STARRS1:2017jku}. When we generate the mock FRB data, the fiducial values of cosmological parameters are taken to be the same with the CMB+BAO+SN results. We use the modified {\tt CAMB} \cite{Lewis:1999bs} and {\tt cosmomc} codes \cite{Lewis:2002ah} with the inclusion of the ePPF module and obtain the posterior probability distribution of cosmological parameters.

\section{Results and discussion} \label{sec:Result}

\subsection{Constraining cosmological parameters} \label{sec:cosmo}

\begin{table}
\renewcommand{\arraystretch}{1.5}
\centering
\caption{Absolute errors ($1\sigma$) of the cosmological parameters in each IDE model from the CMB, FRB1, CMB+FRB1, CBS, FRB2, and CBS+FRB2 data, {assuming fixed DM parameters ($\mu$ and $\gamma$)}. FRB1 and FRB2 denote the FRB data with the event number $N_{\rm FRB}=10^5$ and $N_{\rm FRB}=10^6$, respectively. Here, CBS stands for the CMB+BAO+SN data and $H_0$ is in units of \kms.}
\label{tab:full}
\begin{tabular}{cccccccc}
\hline
Model                 & Parameter          & CMB      & FRB1    & CMB+FRB1 & CBS    & FRB2     & CBS+FRB2  \\
\hline
\multirow{3}{*}{IDE1} & $\beta$       & $0.0025$ & $0.073$ & $0.00084$   & $0.0013$ & $0.023$  & $0.00051$  \\
                      & $H_0$            & $1.8$  & $13$  & $0.34$   & $0.69$ & $10$   & $0.15$  \\
                      & $\Omega_{\rm m}$ & $0.024$  & $0.016$ & 0.0043   & 0.0083 & $0.0049$ & $0.0016$  \\
\hline
\multirow{3}{*}{IDE2 } & $\beta$        & $0.22$   & $0.12$  & $0.022$    & $0.044$  & $0.039$  & $0.020$   \\ 
                      & $H_0$            & $4.9$  & $13$  & $0.31$   & $0.83$ & $13$   & $0.13$  \\
                      & $\Omega_{\rm m}$ & $0.15$   & $0.019$ & $0.0060$   & $0.016$  & $0.0063$ & $0.0037$  \\
\hline
\multirow{3}{*}{IDE3 } & $\beta$        & $0.47$   & $0.16$  & 0.070    & 0.10   & $0.049$  & $0.040$   \\ 
                      & $H_0$            & $4.6$  & $13$  & 0.40   & 0.83 & 13     & $0.16$  \\
                      & $\Omega_{\rm m}$ & $0.15$  & $0.035$ & 0.017    & 0.028  & $0.011$  & $0.0090$  \\
\hline
\multirow{3}{*}{IDE4} & $\beta$        & $0.50$   & $0.23$  & $0.13$     & $0.16$   & $0.071$  & $0.061$   \\
                      & $H_0$           & $3.2$  & $13$  & $0.48$   & $0.82$ & 13     & $0.18$  \\
                      & $\Omega_{\rm m}$ & $0.13$   & $0.043$ & $0.027$    & $0.035$  & $0.014$  & $0.012$   \\
\hline
\end{tabular}
\end{table}

In this section, we study the capability of the mock FRB data of constraining the parameters in the IDE models, with fixed DM parameters. The 1$\sigma$  errors of cosmological parameters are listed in table~\ref{tab:full}. We use FRB1 and FRB2 to denote the mock FRB data in the conservative expectation case (i.e., $N_{\rm FRB}=10^5$) and the optimistic expectation case (i.e., $N_{\rm FRB}=10^6$), respectively.

\begin{figure*}[htb]
{\includegraphics[width=0.45\linewidth]{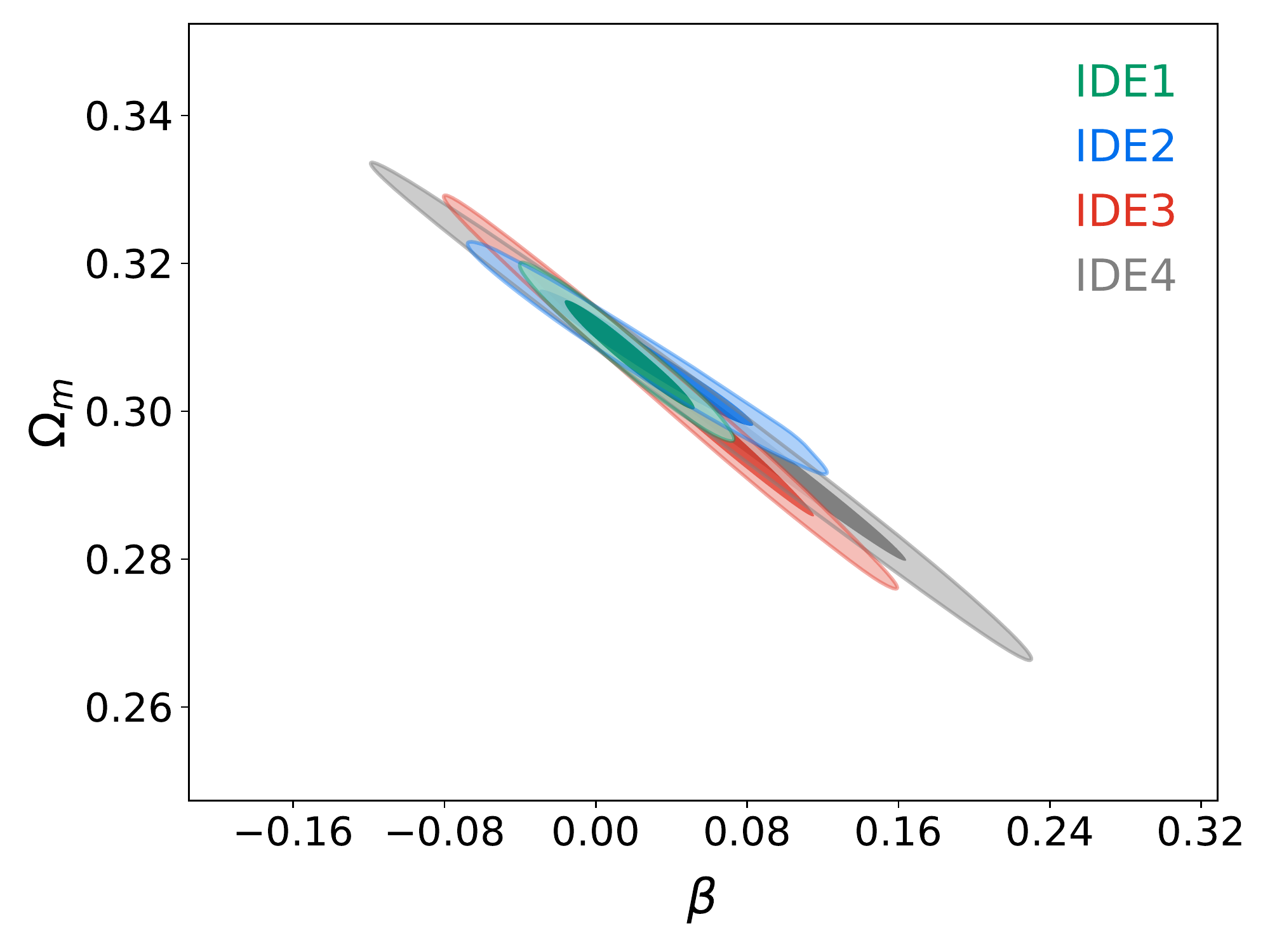}
}
{\includegraphics[width=0.45\linewidth]{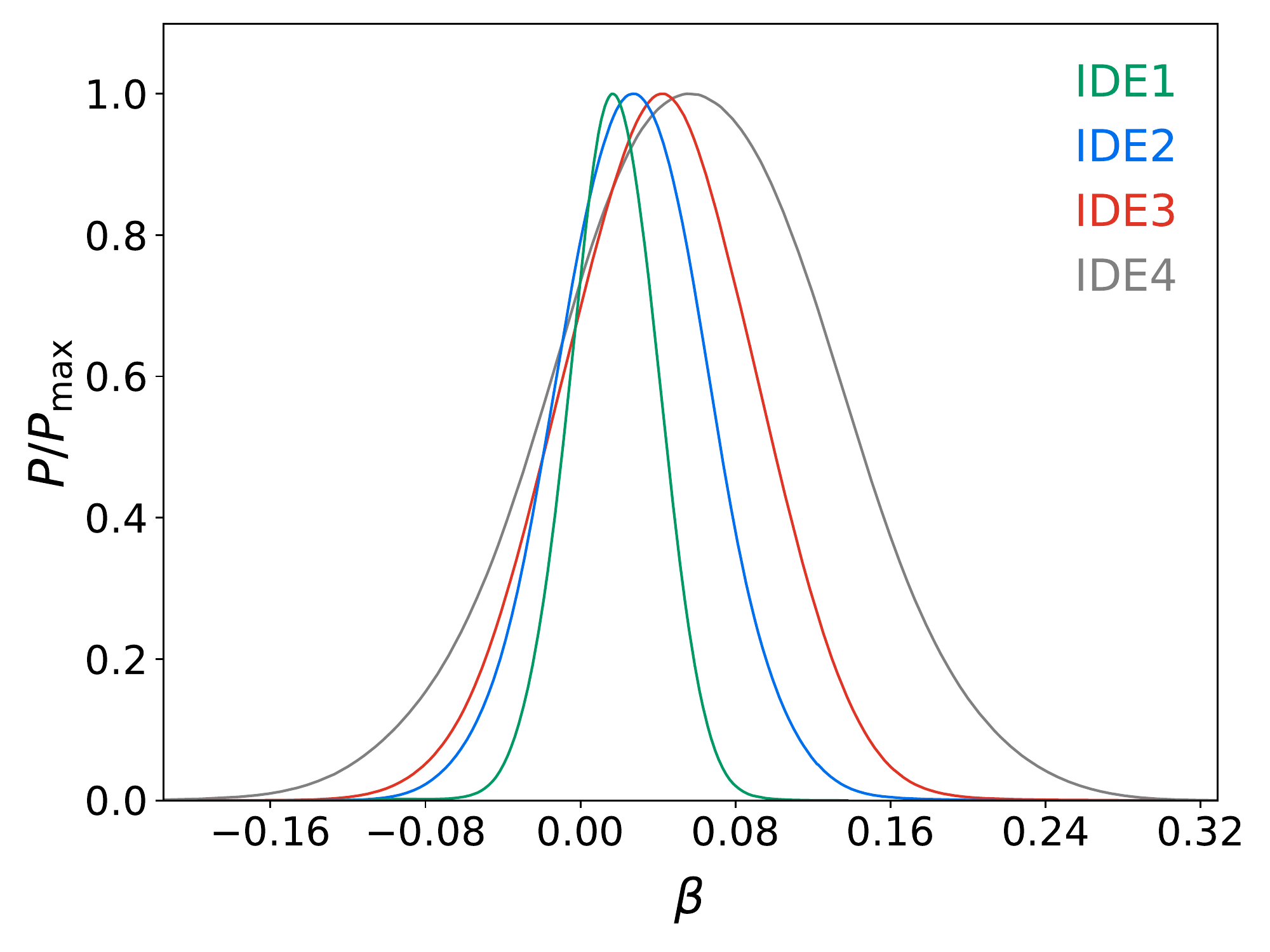}
}
\centering
\caption{Constraints (68.3\% and 95.4\% confidence level) on $\beta$ and $\Omega_{\rm m}$ (left panel) and one-dimensional marginalized probability distributions of $\beta$ (right panel) from the FRB2 data, in each IDE model.
}
 \label{allmodels}
\end{figure*}

First, in all the IDE models, about $10^6$ FRB data can constrain $\beta$ to less than 0.10. We show the constraints from the FRB2 data in each IDE model in figure \ref{allmodels}. Future FRBs provide a way to precisely measure $\beta$ by using only one cosmological probe. In the following, we will discuss the constraints on $\beta$ in each IDE model in sequence. For the IDE1 model ($Q\propto \beta H \rho_{\rm c}$), the CMB data alone can tightly constrain $\beta$, because $H\rho_{\rm c}$ is sensitive to the early-universe probes. The constraint on $\beta$ from the CMB data, $\sigma(\beta)=0.0025$, is even tighter by an order of magnitude than the one from the FRB2 data, $\sigma(\beta)=0.023$. Nevertheless, the combined analysis could still improve the constraint. The CMB+FRB1 data could give $\sigma(\beta)=0.00084$, and the CMB+BAO+SN+FRB2 data could give $\sigma(\beta)=0.00051$. The addition of FRB data could help constrain the absolute error of $\beta$ to achieve less than 0.0010 in the IDE1 model.

\begin{figure*}[htb]
{\includegraphics[width=0.45\linewidth]{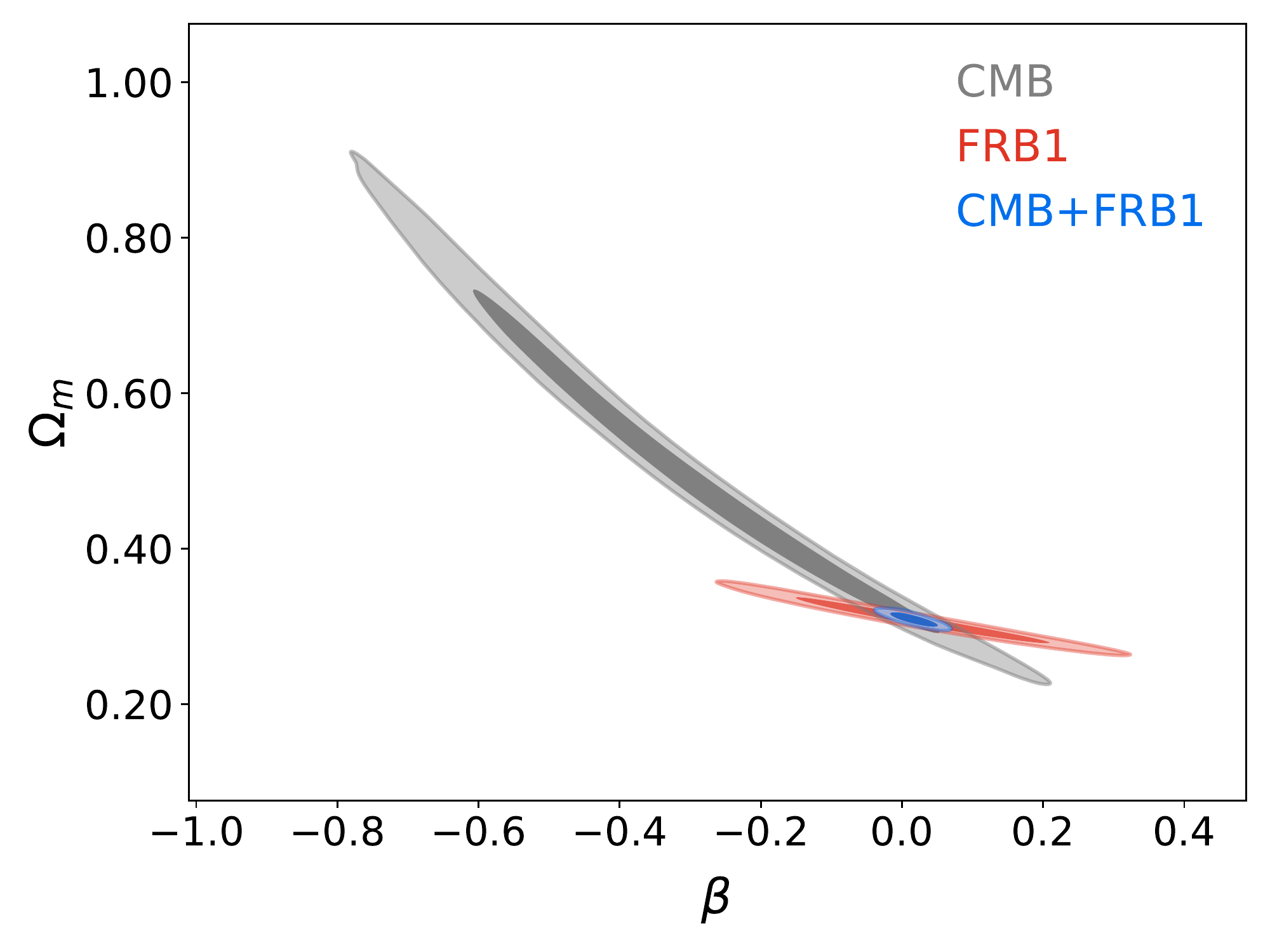}
}
{\includegraphics[width=0.45\linewidth]{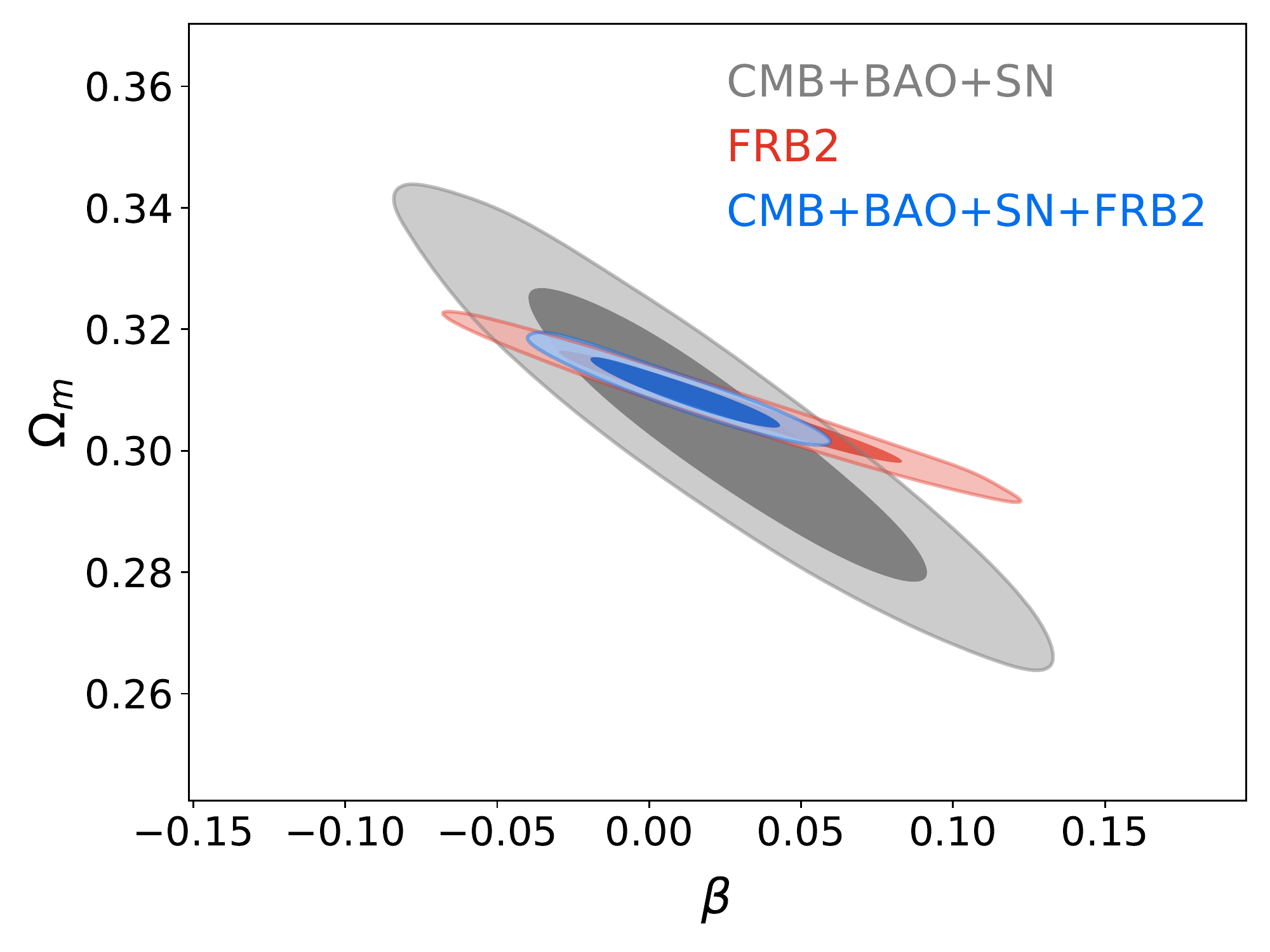}
}
\centering
\caption{Constraints (68.3\% and 95.4\% confidence level) on $\beta$ and $\Omega_{\rm m}$ in the IDE2 model
with $Q=\beta H_0\rho_{\rm c}$ from the CMB, FRB1, and CMB+FRB1 data (left panel) and from the CMB+BAO+SN, FRB2, and CMB+BAO+SN+FRB2 data (right panel).
}
 \label{H0rc}
\end{figure*}

Except for the IDE1 model, the CMB data alone cannot provide tight constraints on $\beta$ in other IDE models. For the IDE2 model, the CMB data can give $\sigma(\beta)=0.22$, and the FRB1 data can give a better constraint with $\sigma(\beta)=0.12$. However, the combined constraint from CMB+FRB1 is significantly better than the result from either the CMB data or the FRB1 data alone. The CMB+FRB1 data can give $\sigma(\beta)=0.022$. Compared to using the CMB data alone and the FRB1 data alone, this combination improves the constraints by about 90\% and 82\%, respectively. The two-dimensional marginalized posterior probability contours are shown in the left panel of figure \ref{H0rc}, from the CMB, FRB1, and CMB+FRB1 data. The orientations of the contours in the $\beta$--$\Omega_{\rm m}$ plane constrained by CMB and by FRB are rather different. Combining the CMB and FRB data could break the parameter degeneracies, and thus evidently improve the constraints. This effect also exists when combining the CMB+BAO+SN and FRB2 data and we show it in the right panel of figure \ref{H0rc}. The addition of the FRB2 data will tighten the CMB+BAO+SN constraint with the absolute error improved from $\sigma(\beta)=0.044$ to $\sigma(\beta)=0.020$.

\begin{figure*}[htb]
{\includegraphics[width=0.45\linewidth]{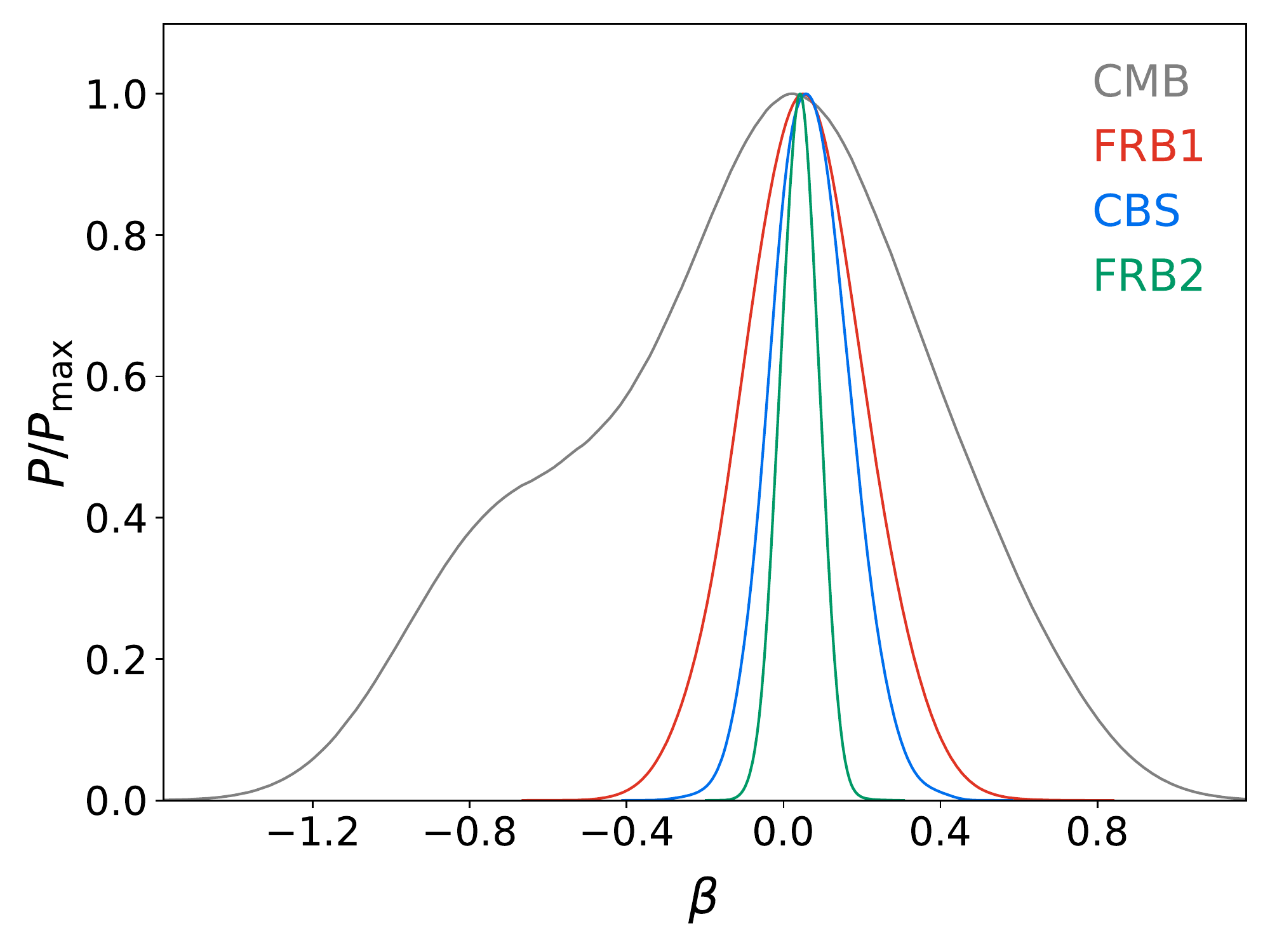}
}
{\includegraphics[width=0.45\linewidth]{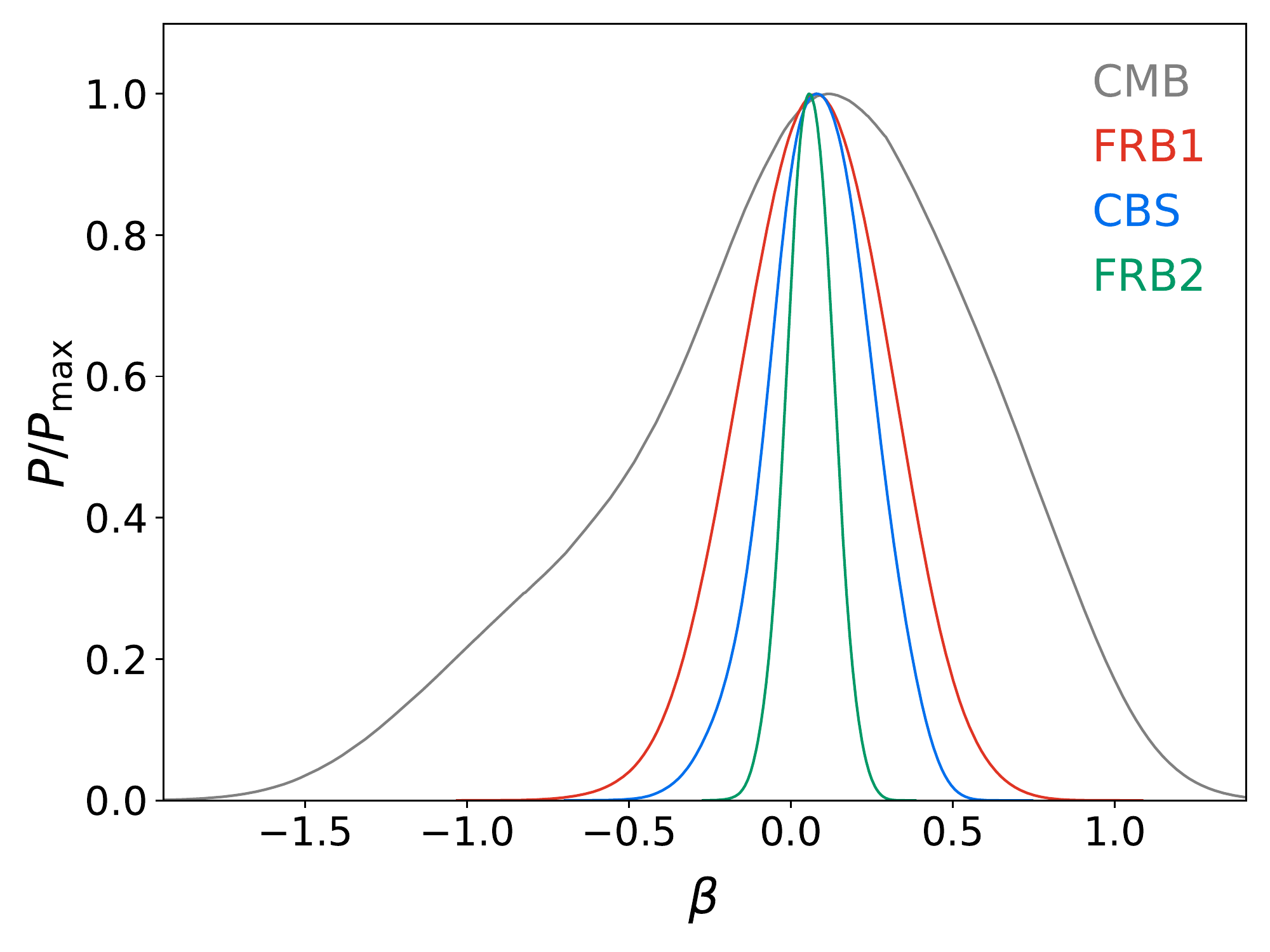}
}
\centering
\caption{One-dimensional marginalized probability distributions of $\beta$ in the IDE3 model with $Q=\beta H\rho_{\rm de}$ (left panel) and in the IDE4 model with $Q=\beta H_0\rho_{\rm de}$ (right panel), from the CMB, FRB1, CMB+BAO+SN, and FRB2 data.
}
 \label{H0rde}
\end{figure*}

For the IDE3 and IDE4 models ($Q\propto \beta \rho_{\rm de}$), even the current mainstream data CMB+BAO+SN cannot constrain the absolute errors of $\beta$ to less than 0.10, urging a new probe to tightly constrain them. For the IDE3 model, the constraint errors on $\beta$ from the CMB, FRB1, CMB+BAO+SN, and FRB2 data are 0.47, 0.16, 0.10, and 0.049, respectively. We plot the comparison of these constraints in the left panel of figure \ref{H0rde}. The CMB+FRB1 data can constrain $\beta$ to less than 0.10, with $\sigma(\beta)=0.070$. The CMB+BAO+SN+FRB2 data can give constraint $\sigma(\beta)=0.040$, slightly tighter than the result from the FRB2 data.

For the IDE4 model, the constraint errors of $\beta$ from the CMB and CMB+BAO+SN data are 0.50 and 0.16, respectively, which are the worst among all the IDE models. The constraints on $\beta$ from the FRB1 data and the FRB2 data are $0.23$ and $0.071$, respectively. The comparison of these constraints is shown in the right panel of figure \ref{H0rde}. We find that, for the IDE models with $Q$ proportional to $\beta \rho_{\rm de}$, about $10^5$ FRB data can give constraint on $\beta$ tighter than the CMB data, but weaker than the CMB+BAO+SN data. About $10^6$ FRB data can give constraint on $\beta$ tighter than the CMB+BAO+SN data. This also displays the advantage of the future FRB data in constraining $\beta$ by one probe. Similar to the results in the IDE3 model, the constraint from the CMB+BAO+SN+FRB2 data, $\sigma(\beta)=0.061$, is only slightly tighter than the result from the FRB2 data. For the IDE models with $Q$ proportional to $\beta \rho_{\rm de}$, the main contribution to the $\beta$ constraints in the combined analysis is from the FRB data.

\begin{figure*}[htb]
{\includegraphics[width=0.45\linewidth]{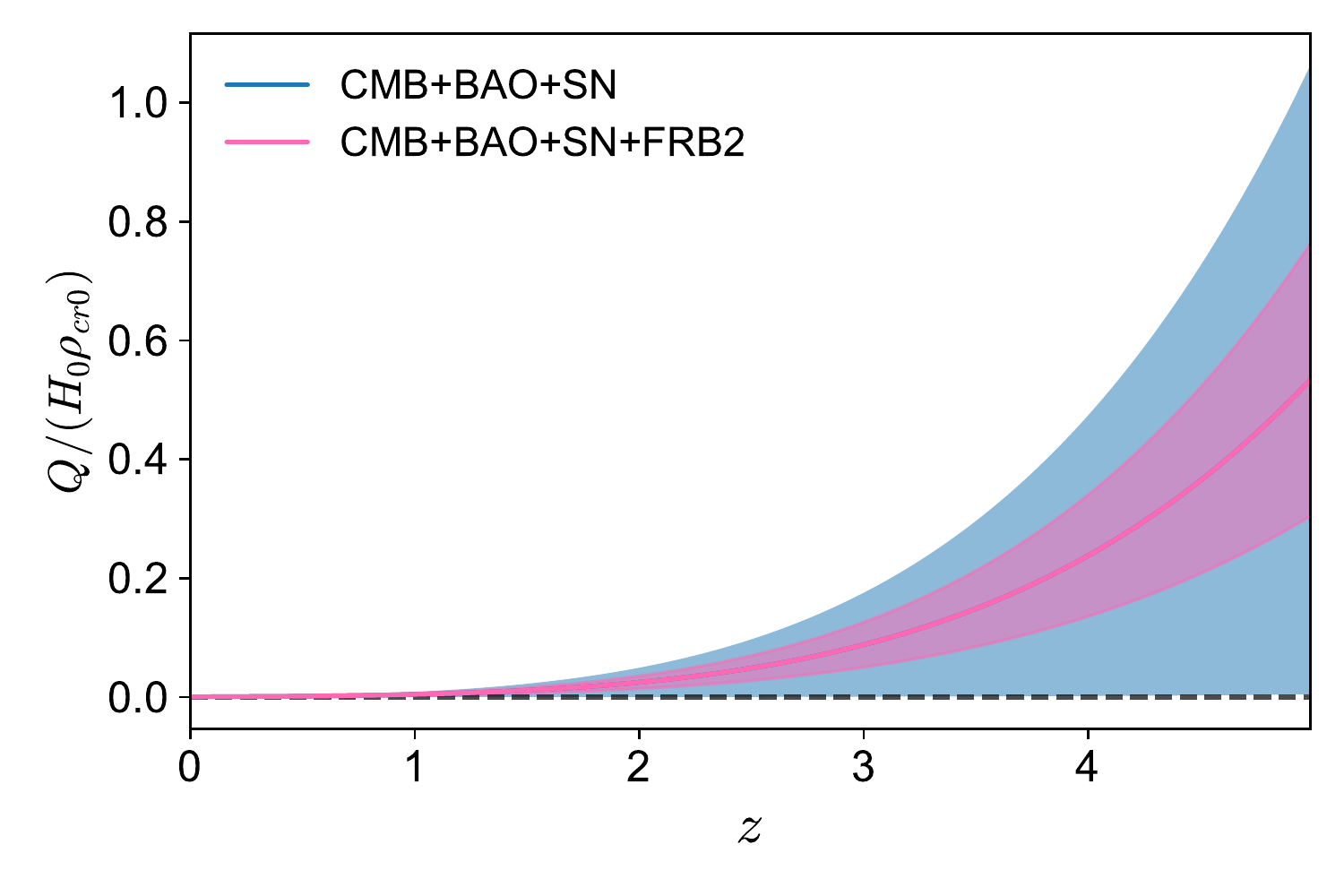}
}
{\includegraphics[width=0.45\linewidth]{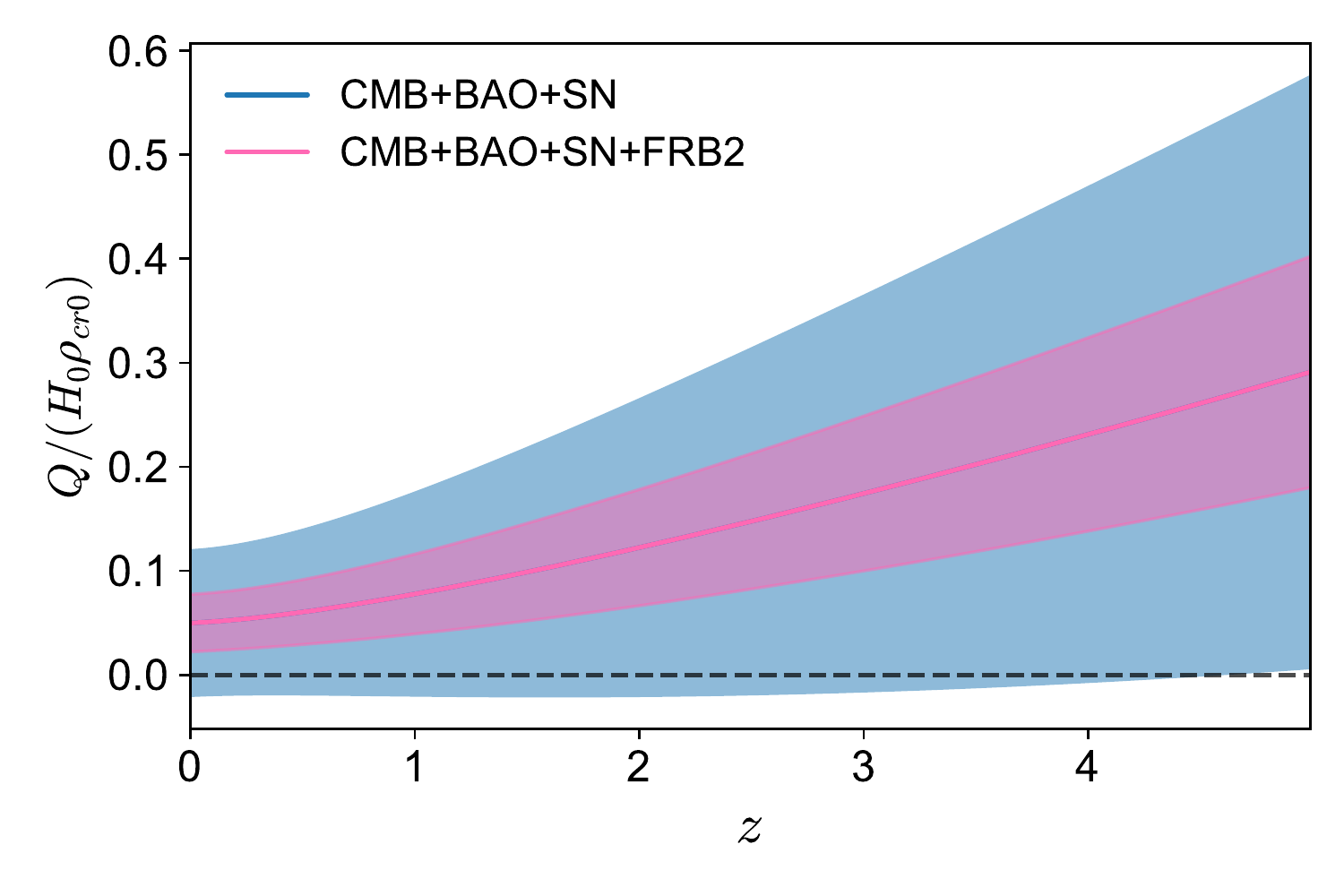}
}
\centering
\caption{The redshift evolutions of  $Q/(H_0 \rho_{\rm cr0})$ in the IDE1 model with $Q=\beta H\rho_{\rm c}$ (left panel)
and in the IDE3 model with $Q=\beta H\rho_{\rm de}$ (right panel). The blue and pink shaded regions represent the $1\sigma$ constraints from the CMB+BAO+SN and CMB+BAO+SN+FRB2 data, respectively. The black dashed lines denote the $\Lambda$CDM model ($Q=0$).
}
 \label{recons}
\end{figure*}

The redshift evolutions of the interaction term $Q(z)$ can be reconstructed by the constraints. In figure \ref{recons}, we show the evolution of $Q/(H_0 \rho_{\rm cr0})$ versus $z$ in the IDE1 and IDE3 models, where $\rho_{\rm cr0}=3 M_{\rm pl}^2 H_0^2$ is the present-day critical density of the Universe. The best-fit line and the $1\sigma$ region are shown in the figure. We find that in the IDE1 model, although the CMB+BAO+SN data alone can precisely constrain $\beta$, the addition of the FRB data could significantly shrink the uncertainty of $Q$ in the redshift region with $z>3$. In the IDE3 model, the addition of the FRB data could help constrain the evolution of $Q$ in the whole redshift region.

\begin{figure*}[htb]
{\includegraphics[width=0.45\linewidth]{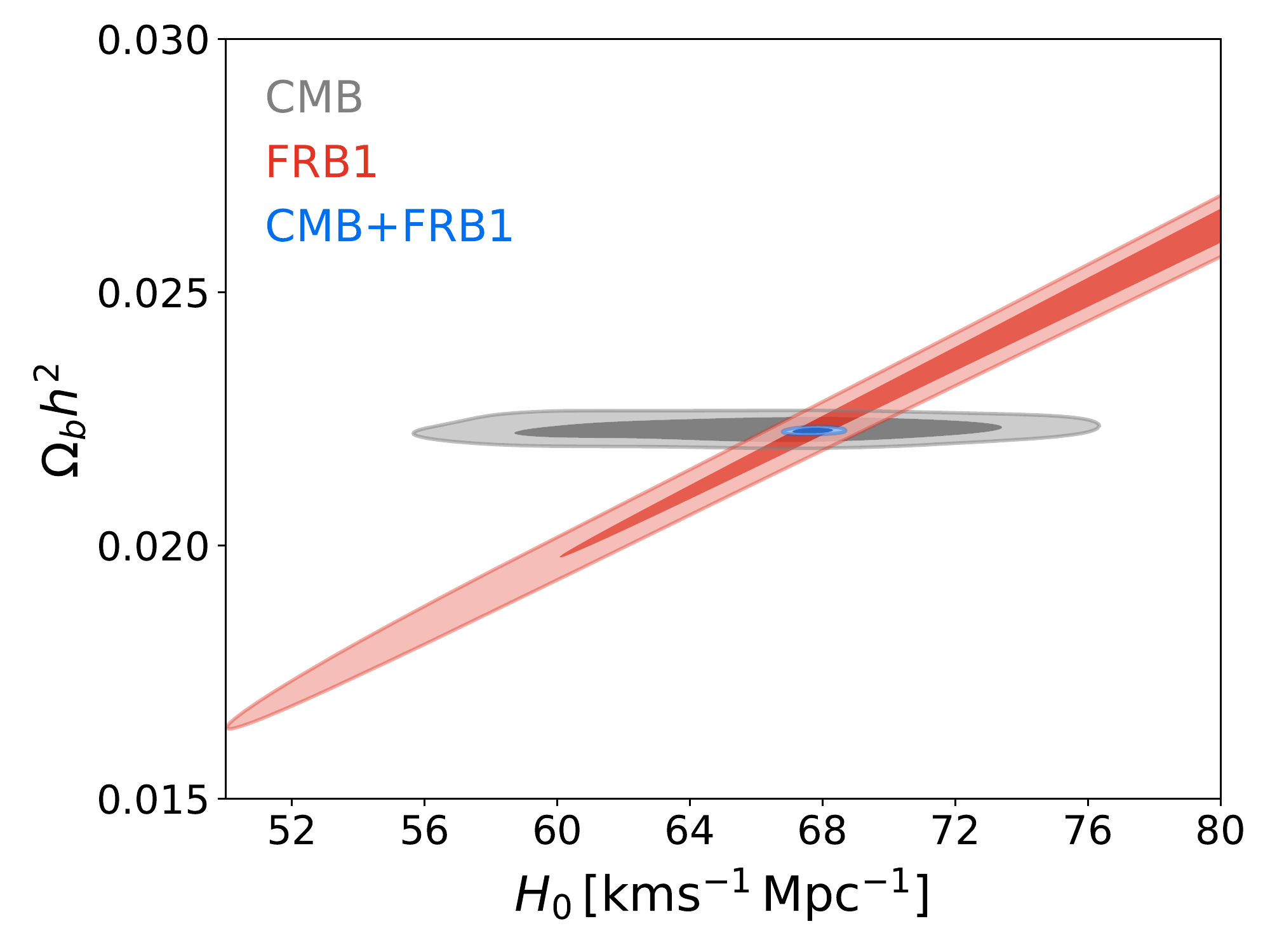}}
{\includegraphics[width=0.45\linewidth]{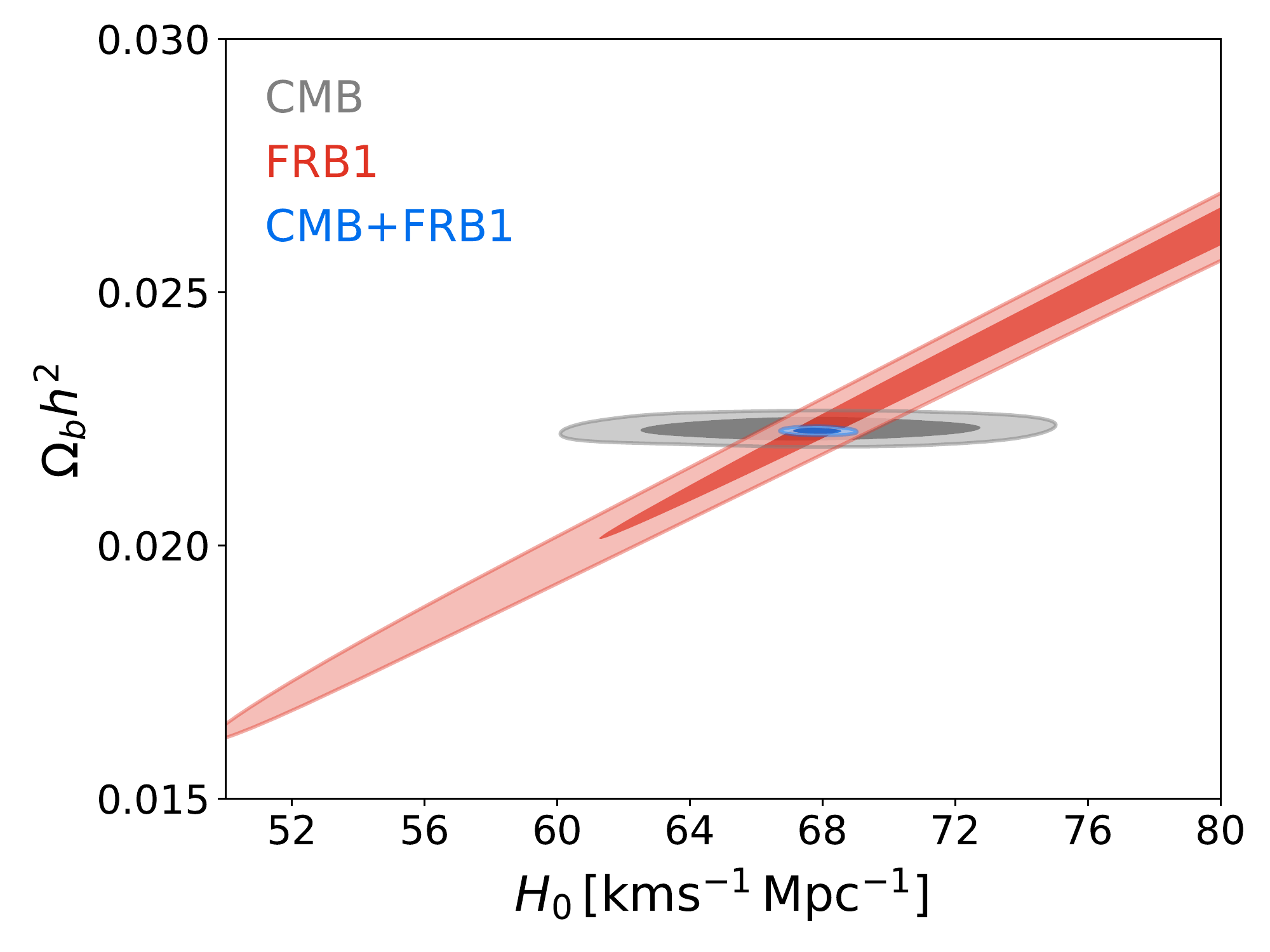}
}
\centering
\caption{Constraints (68.3\% and 95.4\% confidence level) on $H_0$ and $\Omega_{\rm b}h^2$ in the IDE3 model with $Q=\beta H\rho_{\rm de}$ (left panel) and the IDE4 model with $Q=\beta H_0\rho_{\rm de}$ (right panel) from the CMB, FRB1, and CMB+FRB1 data.
}
 \label{H0}
\end{figure*}

We briefly discuss the constraints on $H_0$. The FRB data alone cannot effectively constrain $H_0$, since DM is proportional to $\Omega_{\rm b}H_0$. However, the CMB data could provide tight constraint on $\Omega_{\rm b}H_0^2$, resulting in tight $H_0$ constraints if combining the CMB and FRB data. This effect can be seen in figure \ref{H0}, in accordance with the previous experience in the $w$CDM and CPL models \cite{Zhao:2020ole}. In all the IDE models, even the CMB+BAO+SN data cannot constrain $H_0$ at sub-percent precision, yet the CMB+FRB1 data could do. For example, in the IDE4 model, the CMB data can only provide a 4.7\% measurement for $H_0$, while the CMB+FRB1 data can constrain $H_0$ to a 0.71\% precision. The constraint is improved by about 85\%. The CMB+BAO+SN data provide a 1.2\% measurement for $H_0$, whereas the combined CMB+BAO+SN+FRB2 data provide a 0.27\% measurement.

\begin{figure*}[htbp]
{\includegraphics[width=0.6\linewidth]{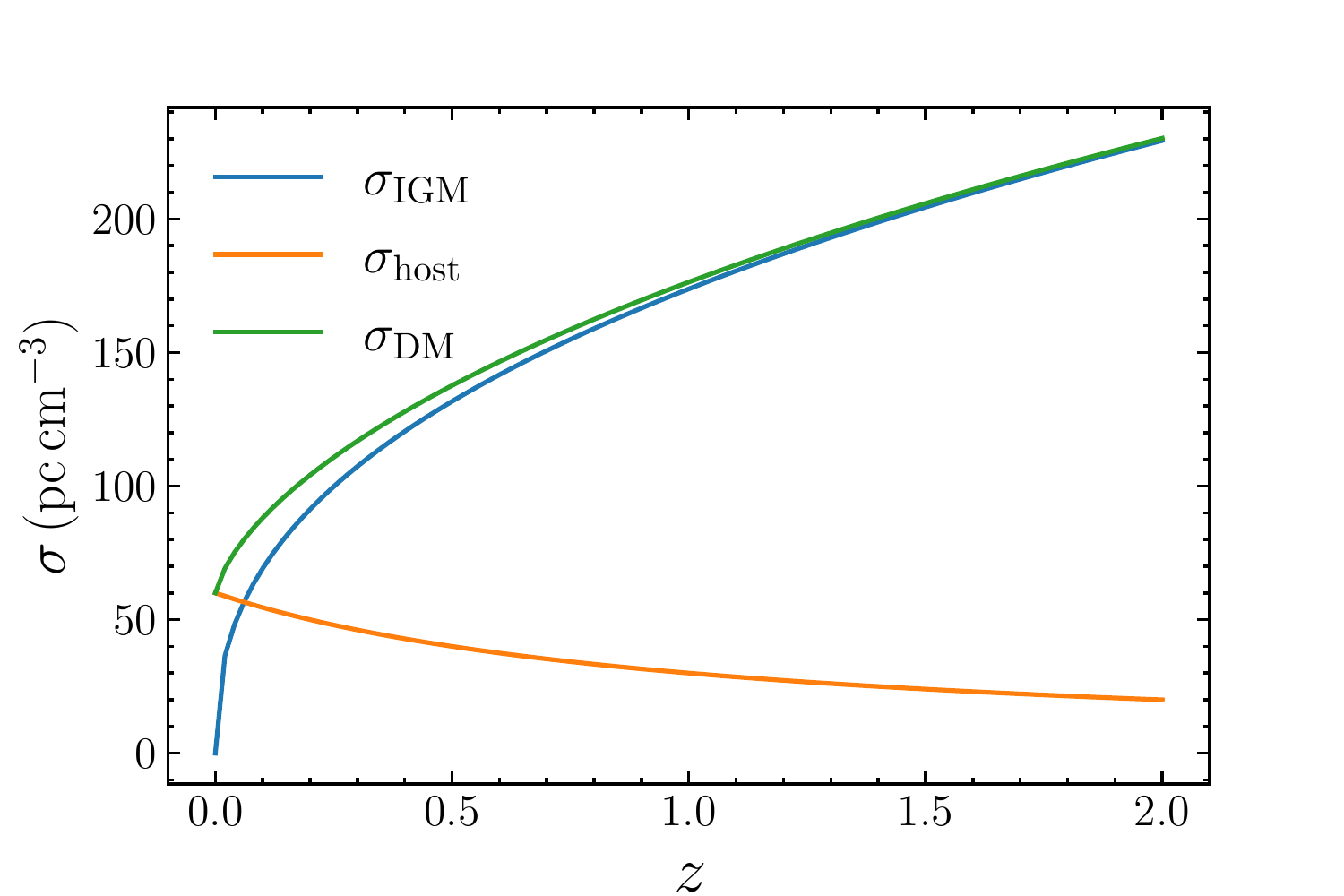}
}
\centering
\caption{ The evolutions of $\sigma_{\rm IGM}$, $\sigma_{\rm host}$, and $\sigma_{\rm{DM}}$, assuming $\sigma_{\rm host}=60$\pccc.}
 \label{sig}
\end{figure*}

The values of $\sigma_{\rm host}$ have minor effect on the cosmological constraints. We calculate two new cases for the IDE3 model assuming $\sigma_{\rm host}=20$\pccc and $\sigma_{\rm host}=60$\pccc. For example, $10^6$ FRB data with $\sigma_{\rm host}=60$\pccc could give $\sigma(\beta)=0.051$ in the IDE3 model, while $10^6$ FRB data with $\sigma_{\rm host}=20$\pccc could give $\sigma(\beta)=0.049$. There are two reasons: (i) the values of $\sigma_{\rm IGM}$ are larger than $\sigma_{\rm host}$, and (ii) $\sigma_{\rm host}$ contributes to the total uncertainties by multiplying a factor $1/(1+z)$. In figure \ref{sig}, we can clearly see that $\sigma_{\rm{DM}}$ is dominated by $\sigma_{\rm IGM}$ except for very low redshift.

\subsection{Jointly constraining cosmological parameters and DM parameters} \label{sec:join}

\begin{table*}[htb]
\renewcommand{\arraystretch}{1.5}
\caption{Constraints on the IDE1 and IDE3 models and the DM parameters from $10^5$ mock FRB data (FRB1) and $10^6$ mock FRB data (FRB2), with different fixed DM parameters. Here, $\mu$ is in units of \pccc.}
\centering
\vbox{}
\begin{tabular}{cccccccccc}
\toprule
           \multirow{2}{*}{ Sample}      &   \multirow{2}{*}{\begin{tabular}[c]{@{}c@{}}DM parameter\end{tabular}}       & \multicolumn{4}{c}{IDE1\,($Q=\beta H\rho_{\rm c}$)}                                                                   & \multicolumn{4}{c}{IDE3\,($Q=\beta H\rho_{\rm de}$)}                                                                  \\ \cmidrule(lr){3-6} \cmidrule(lr){7-10}
                      &  & \multicolumn{1}{c}{$\sigma(\beta)$}  & \multicolumn{1}{c}{$\sigma(\Omega_{\rm m})$}     & \multicolumn{1}{c}{$\sigma(\mu)$}  & $\sigma(\gamma)$ & \multicolumn{1}{c}{$\sigma(\beta)$}  & \multicolumn{1}{c}{$\sigma(\Omega_{\rm m})$}    & \multicolumn{1}{c}{$\sigma(\mu)$}  & $\sigma(\gamma)$ \\
\midrule
\multirow{3}{*}{FRB1} & $\mu$, $\gamma$ fixed & \multicolumn{1}{c}{0.073}  & \multicolumn{1}{c}{0.016}  & \multicolumn{1}{c}{ } &   & \multicolumn{1}{c}{0.16}  & \multicolumn{1}{c}{0.035} & \multicolumn{1}{c}{ } &   \\ 

                      & $\gamma$ fixed  & \multicolumn{1}{c}{0.10}  & \multicolumn{1}{c}{0.028}  & \multicolumn{1}{c}{3.6} &   & \multicolumn{1}{c}{0.26}  & \multicolumn{1}{c}{0.068} & \multicolumn{1}{c}{4.4} &   \\ 
                      & both free & \multicolumn{1}{c}{0.15}  & \multicolumn{1}{c}{0.091}  & \multicolumn{1}{c}{4.9}   & 1.1  & \multicolumn{1}{c}{0.50}  & \multicolumn{1}{c}{0.16}  & \multicolumn{1}{c}{5.2}   & 1.2  \\
\midrule
\multirow{3}{*}{FRB2} & $\mu$, $\gamma$ fixed & \multicolumn{1}{c}{0.023} & \multicolumn{1}{c}{0.0049} & \multicolumn{1}{c}{ } &   & \multicolumn{1}{c}{0.049} & \multicolumn{1}{c}{0.011} & \multicolumn{1}{c}{ } &   \\ 

                      & $\gamma$ fixed   & \multicolumn{1}{c}{0.033} & \multicolumn{1}{c}{0.0087} & \multicolumn{1}{c}{1.1} &   & \multicolumn{1}{c}{0.085} & \multicolumn{1}{c}{0.021} & \multicolumn{1}{c}{1.4} &   \\ 
                      & both free  & \multicolumn{1}{c}{0.050} & \multicolumn{1}{c}{0.041}  & \multicolumn{1}{c}{1.7}  & 0.48 & \multicolumn{1}{c}{0.137}  & \multicolumn{1}{c}{0.056} & \multicolumn{1}{c}{1.5}  & 0.64 \\
\bottomrule
\end{tabular}\label{free}
\end{table*}

\begin{figure*}[htbp]
{\includegraphics[width=0.45\linewidth]{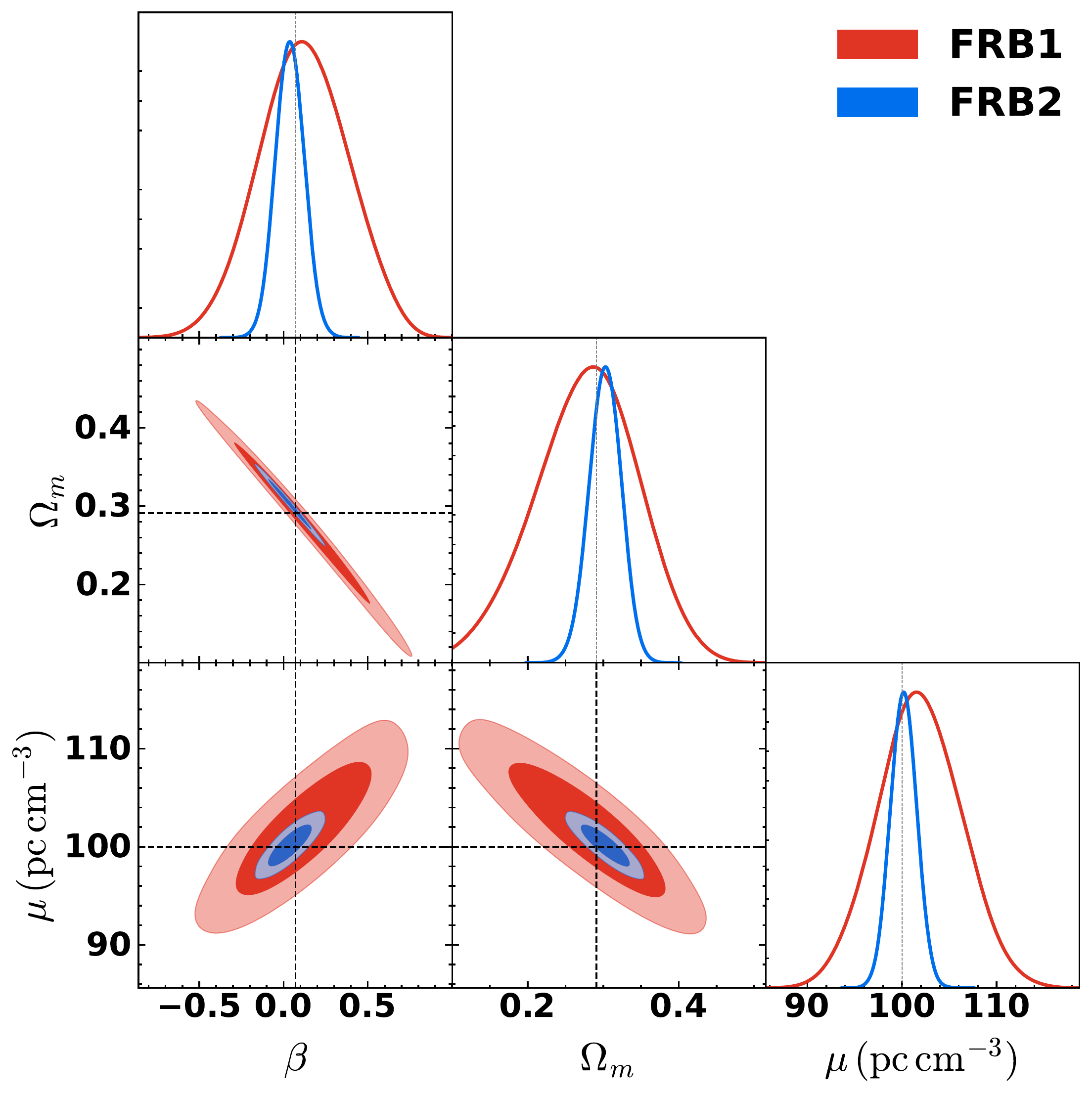}
}
{\includegraphics[width=0.45\linewidth]{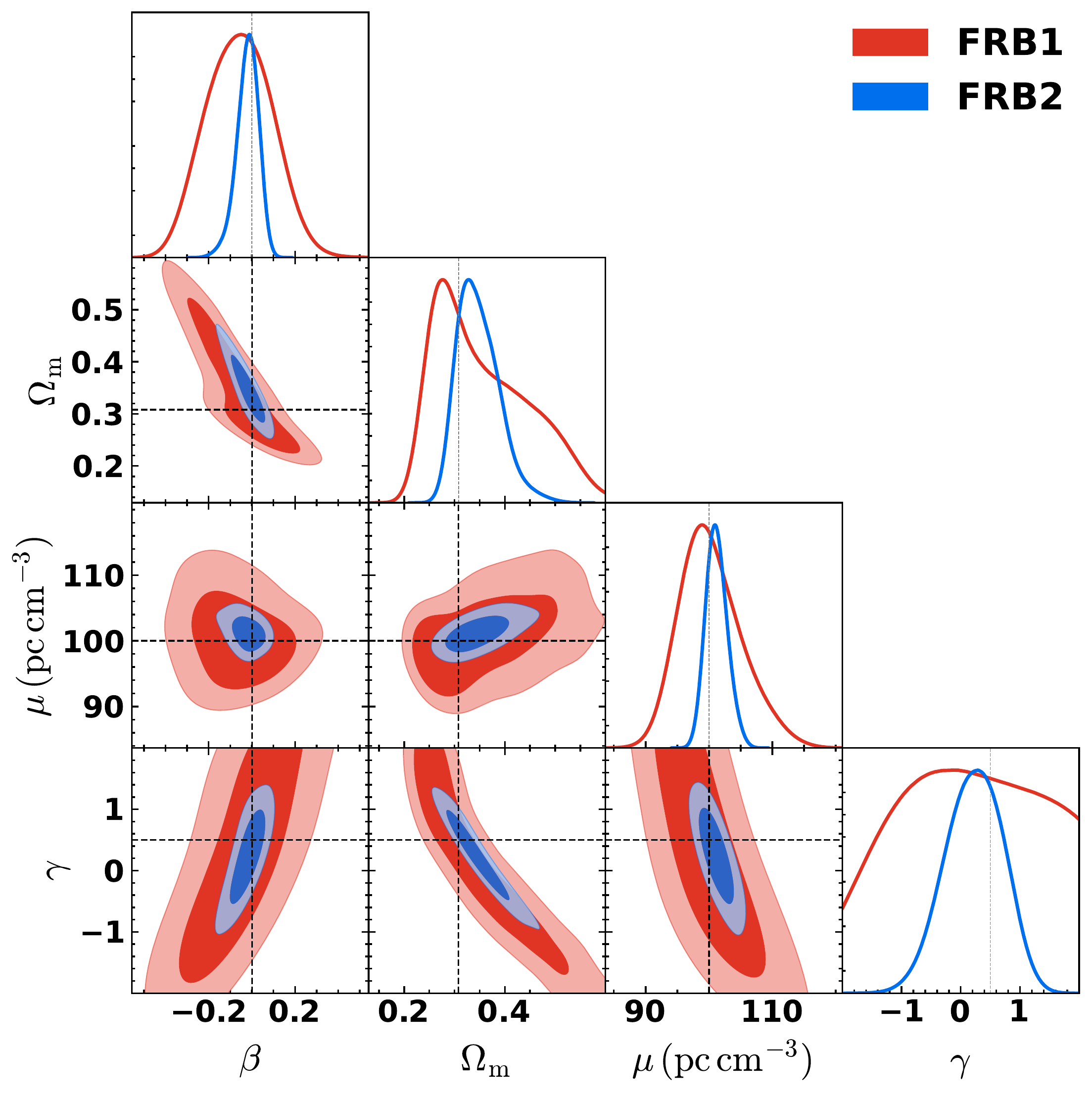}
}
\centering
\caption{Constraints on the IDE3 model ($Q=\beta H\rho_{\rm de}$) and the mean value of $\DM_{\rm host}$ (left panel) and the IDE1 model ($Q=\beta H\rho_{\rm c}$) and the mean value and redshift evolution of $\DM_{\rm host}$ (right panel), using the FRB1 and FRB2 data. The fiducial values are denoted by dotted lines.}
\label{Hrde}
\end{figure*}

We use the IDE1 and IDE3 models as examples to illustrate the effect of treating the DM parameters as free parameters. The constraints on the IDE1 and IDE3 models and the DM parameters are listed in table \ref{free}. We first consider the case without $\DM_{\rm host}$ redshift evolution (i.e., $\gamma=0$). The constraint contours for the IDE3 model from the FRB1 and FRB2 data are shown in the left panel of figure \ref{Hrde}, and the fiducial values are denoted by dotted lines. We can clearly see that all constraints of the DM parameters and cosmological parameters could cover their fiducial values.

Because of the correlations between $\mu$ and cosmological parameters, treating $\mu$ as a free parameter weakens the constraints on $\beta$ and $\Omega_{\rm m}$ in the global fit. If $\mu$ is free, the absolute errors of $\beta$ are broader by about 50\%, compared to the ones with fixed $\mu$. For example, the FRB1 data can give $\sigma(\beta)=0.16$ in the IDE3 model with fixed $\mu$, and give  $\sigma(\beta)=0.26$ when $\mu$ is a free parameter, which is 62.5\% looser than the former constraint. The constraints of $\beta$ from the FRB1 and FRB2 data are still comparable to the CMB and CMB+BAO+SN results, respectively. The main results do not change significantly.

For the case that $\mu$ and $\gamma$ both are free, the constraints on the IDE1 model are shown in the right panel of figure \ref{Hrde} and can also cover the fiducial values. The absolute errors of $\beta$ are about 1--2 times broader than the ones with fixed $\mu$ and $\gamma$. For example, in the IDE1 model, the constraint $\sigma(\beta)=0.15$ from the FRB1 data when $\mu$ and $\gamma$ both are free parameters is 105.5\% looser than the constraint with fixed $\mu$ and $\gamma$, $\sigma(\beta)=0.073$.

{The constraint errors of the DM parameters are similar for different IDE models. For example, in the case of $\gamma$ fixed, the FRB2 data could give $\sigma(\mu)= 1.1$\pccc and $\sigma(\mu)= 1.4$\pccc for the IDE1 model and the IDE3 model, respectively. In all the cases (i.e., in both IDE1 and IDE3 models with whether $\gamma$ fixed or not), the FRB1 data could give $\sigma(\mu)\sim 4.5$\pccc and the FRB2 data could give $\sigma(\mu)\sim 1.5$\pccc. However, we note that these constraints are somewhat meaningless, because with a large number of FRB data, the asymmetry of $\DM_{\rm host}$ distribution would affect the constraints. Using a log-normal distribution may be more appropriate to account for observed high $\DM_{\rm host}$ values \cite{Macquart:2020lln,KochOcker:2022ook}. In this work, we focus on the constraints on cosmological parameters, and leave the detailed research on $\DM_{\rm host}$ distribution to future work.} The redshift evolution of ${\rm DM}_{\rm host}$, $\gamma$, is hard to be constrained by the FRB1 data, but the FRB2 data could constrain its absolute error to about 0.60. However, we still need more data or new methods to precisely constrain it.

\section{Conclusion} \label{sec:con}

In this paper, we study how many FRBs are needed to precisely measure the interaction between dark energy and cold dark matter, by constraining $\beta$ in the IDE models. Four phenomenological IDE models are considered, i.e., $Q=\beta H\rho_{\rm c}$ (IDE1), $Q=\beta H_0\rho_{\rm c}$ (IDE2), $Q=\beta H\rho_{\rm de}$ (IDE3), and $Q=\beta H_0\rho_{\rm de}$ (IDE4). We simulate the future FRB data with data number $N_{\rm FRB}=10^5$ (FRB1) and $N_{\rm FRB}=10^6$ (FRB2), based on the current knowledge of FRBs from the first CHIME/FRB catalog and ASKAP.

We find that, if we fix the FRB properties, in all the IDE models, about $10^6$ FRB data alone can constrain the absolute error of $\beta$ to less than $0.10$. This provides a way to constrain $\beta$ by only one cosmological probe. For the IDE1 model, although the CMB data can provide tight constraint on $\beta$, the addition of the FRB data could still improve the constraint. The combined analysis is also important in the IDE2 model. In this model, the CMB and FRB data could obviously break the parameter degeneracies inherent in each other. Therefore, compared to the result from the CMB data or the CMB+BAO+SN data, the combined constraint on $\beta$ is significantly improved with the addition of the FRB data.

For the IDE3 and IDE4 models with $Q$ proportional to $\beta \rho_{\rm de}$, even the current CMB+BAO+SN data cannot constrain the absolute errors of $\beta$ to less than $0.10$. We find that about $10^5$ FRB data can give constraint on $\beta$ tighter than the CMB data and about $10^6$ FRB data can give constraint on $\beta$ tighter than the CMB+BAO+SN data. In the combined analysis, the main contribution to the combined constraints is from the FRB data. As a low-redshift probe, the FRB data alone could play an important role in these models.

We reconstruct the redshift evolution of the interaction term $Q(z)$. The addition of the FRB data could help check whether the interaction exists at a high confidence level, and eventually provide evidence to distinguish between the $\Lambda$CDM model and the IDE models. With respect to the Hubble constant, the CMB+FRB1 data could constrain $H_0$ to 1\% precision in all the IDE models, which is tighter than the corresponding results from the CMB+BAO+SN data.

We also treat the DM parameters as free parameters to study the effects of the correlations between them and cosmological parameters. Jointly constraining the FRB
properties and cosmological parameters would increase the constraint errors of $\beta$ by a factor of about 0.5--2. In this work, we use the Macquart relation to constrain the IDE models at the background level. A future work could use auto-correlations of FRBs and cross-correlations with the large-scale structure to constrain cosmological parameters with clustering effects \cite{Masui:2015ola,Reischke:2020cgd,Shirasaki:2021rzq,Rafiei-Ravandi:2021hbw}.

\acknowledgments
We thank Jing-Zhao Qi, Hai-Li Li, and Yichao Li for helpful discussions. This work was supported by the National SKA Program of China (Grants Nos. 2022SKA0110200 and 2022SKA0110203) and the National Natural
Science Foundation of China (Grants Nos. 11975072,
11835009, and 11875102).

\bibliography{FRBIDE}

\providecommand{\href}[2]{#2}\begingroup\raggedright\begin{thebibliography}{100}

\bibitem{CHIMEFRB:2020abu}
{\scshape CHIME/FRB} collaboration, \emph{{A bright millisecond-duration radio
  burst from a Galactic magnetar}},
  \href{https://doi.org/10.1038/s41586-020-2863-y}{\emph{Nature} {\bfseries
  587} (2020) 54} [\href{https://arxiv.org/abs/2005.10324}{{\ttfamily
  2005.10324}}].

\bibitem{Bochenek:2020zxn}
C.D.~Bochenek, V.~Ravi, K.V.~Belov, G.~Hallinan, J.~Kocz, S.R.~Kulkarni et~al.,
  \emph{{A fast radio burst associated with a Galactic magnetar}},
  \href{https://doi.org/10.1038/s41586-020-2872-x}{\emph{Nature} {\bfseries
  587} (2020) 59} [\href{https://arxiv.org/abs/2005.10828}{{\ttfamily
  2005.10828}}].

\bibitem{Macquart:2020lln}
J.P.~Macquart et~al., \emph{{A census of baryons in the Universe from localized
  fast radio bursts}},
  \href{https://doi.org/10.1038/s41586-020-2300-2}{\emph{Nature} {\bfseries
  581} (2020) 391} [\href{https://arxiv.org/abs/2005.13161}{{\ttfamily
  2005.13161}}].

\bibitem{Deng:2013aga}
W.~Deng and B.~Zhang, \emph{{Cosmological Implications of Fast Radio
  Burst/Gamma-Ray Burst Associations}},
  \href{https://doi.org/10.1088/2041-8205/783/2/L35}{\emph{Astrophys. J. Lett.}
  {\bfseries 783} (2014) L35}
  [\href{https://arxiv.org/abs/1401.0059}{{\ttfamily 1401.0059}}].

\bibitem{Gao:2014iva}
H.~Gao, Z.~Li and B.~Zhang, \emph{{Fast Radio Burst/Gamma-Ray Burst
  Cosmography}},
  \href{https://doi.org/10.1088/0004-637X/788/2/189}{\emph{Astrophys. J.}
  {\bfseries 788} (2014) 189}
  [\href{https://arxiv.org/abs/1402.2498}{{\ttfamily 1402.2498}}].

\bibitem{Zhou:2014yta}
B.~Zhou, X.~Li, T.~Wang, Y.-Z.~Fan and D.-M.~Wei, \emph{{Fast radio bursts as a
  cosmic probe?}},
  \href{https://doi.org/10.1103/PhysRevD.89.107303}{\emph{Phys. Rev. D}
  {\bfseries 89} (2014) 107303}
  [\href{https://arxiv.org/abs/1401.2927}{{\ttfamily 1401.2927}}].

\bibitem{Yang:2016zbm}
Y.-P.~Yang and B.~Zhang, \emph{{Extracting host galaxy dispersion measure and
  constraining cosmological parameters using fast radio burst data}},
  \href{https://doi.org/10.3847/2041-8205/830/2/L31}{\emph{Astrophys. J. Lett.}
  {\bfseries 830} (2016) L31}
  [\href{https://arxiv.org/abs/1608.08154}{{\ttfamily 1608.08154}}].

\bibitem{Li:2017mek}
Z.-X.~Li, H.~Gao, X.-H.~Ding, G.-J.~Wang and B.~Zhang, \emph{{Strongly lensed
  repeating fast radio bursts as precision probes of the universe}},
  \href{https://doi.org/10.1038/s41467-018-06303-0}{\emph{Nature Commun.}
  {\bfseries 9} (2018) 3833}
  [\href{https://arxiv.org/abs/1708.06357}{{\ttfamily 1708.06357}}].

\bibitem{Walters:2017afr}
A.~Walters, A.~Weltman, B.M.~Gaensler, Y.-Z.~Ma and A.~Witzemann, \emph{{Future
  Cosmological Constraints from Fast Radio Bursts}},
  \href{https://doi.org/10.3847/1538-4357/aaaf6b}{\emph{Astrophys. J.}
  {\bfseries 856} (2018) 65}
  [\href{https://arxiv.org/abs/1711.11277}{{\ttfamily 1711.11277}}].

\bibitem{Jaroszynski:2018vgh}
M.~Jaroszynski, \emph{{Fast Radio Bursts and cosmological tests}},
  \href{https://doi.org/10.1093/mnras/sty3529}{\emph{Mon. Not. Roy. Astron.
  Soc.} {\bfseries 484} (2019) 1637}
  [\href{https://arxiv.org/abs/1812.11936}{{\ttfamily 1812.11936}}].

\bibitem{Liu:2019jka}
B.~Liu, Z.~Li, H.~Gao and Z.-H.~Zhu, \emph{{Prospects of strongly lensed
  repeating fast radio bursts: Complementary constraints on dark energy
  evolution}}, \href{https://doi.org/10.1103/PhysRevD.99.123517}{\emph{Phys.
  Rev. D} {\bfseries 99} (2019) 123517}
  [\href{https://arxiv.org/abs/1907.10488}{{\ttfamily 1907.10488}}].

\bibitem{Liu:2019ddm}
T.~Liu, S.~Cao, J.~Zhang, S.~Geng, Y.~Liu, X.~Ji et~al., \emph{{Implications
  from simulated strong gravitational lensing systems: constraining
  cosmological parameters using Gaussian Processes}},
  \href{https://doi.org/10.3847/1538-4357/ab4bc3}{\emph{Astrophys. J.}
  {\bfseries 886} (2019) 94}
  [\href{https://arxiv.org/abs/1910.02592}{{\ttfamily 1910.02592}}].

\bibitem{Zhang:2020btn}
L.~Zhang and Z.~Li, \emph{{Combinations of Standard Pings and Standard Candles:
  An Effective and Hubble Constant-free Probe of Dark Energy Evolution}},
  \href{https://doi.org/10.3847/1538-4357/abb091}{\emph{Astrophys. J.}
  {\bfseries 901} (2020) 130}.

\bibitem{Qiang:2021bwb}
D.-C.~Qiang and H.~Wei, \emph{{Effect of Redshift Distributions of Fast Radio
  Bursts on Cosmological Constraints}},
  \href{https://doi.org/10.1103/PhysRevD.103.083536}{\emph{Phys. Rev. D}
  {\bfseries 103} (2021) 083536}
  [\href{https://arxiv.org/abs/2102.00579}{{\ttfamily 2102.00579}}].

\bibitem{Dai:2021czy}
J.-P.~Dai and J.-Q.~Xia, \emph{{Reconstruction of baryon fraction in
  intergalactic medium through dispersion measurements of fast radio bursts}},
  \href{https://doi.org/10.1093/mnras/stab785}{\emph{Mon. Not. Roy. Astron.
  Soc.} {\bfseries 503} (2021) 4576}
  [\href{https://arxiv.org/abs/2103.08479}{{\ttfamily 2103.08479}}].

\bibitem{Zhao:2021jeb}
S.~Zhao, B.~Liu, Z.~Li and H.~Gao, \emph{{Model-independent Estimation of H 0
  and \ensuremath{\Omega} K from Strongly Lensed Fast Radio Bursts}},
  \href{https://doi.org/10.3847/1538-4357/abfa91}{\emph{Astrophys. J.}
  {\bfseries 916} (2021) 70}.

\bibitem{Zhu:2022mzv}
C.~Zhu and J.~Zhang, \emph{{Improvement of cosmological constraints with the
  cross-correlation between line-of-sight optical galaxy and FRB dispersion
  measures}}, \href{https://doi.org/10.1103/PhysRevD.106.023513}{\emph{Phys.
  Rev. D} {\bfseries 106} (2022) 023513}
  [\href{https://arxiv.org/abs/2205.03867}{{\ttfamily 2205.03867}}].

\bibitem{Wu:2022dgy}
P.-J.~Wu, Y.~Shao, S.-J.~Jin and X.~Zhang, \emph{{A path to precision
  cosmology: Synergy between four promising late-universe cosmological
  probes}},  \href{https://arxiv.org/abs/2202.09726}{{\ttfamily 2202.09726}}.

\bibitem{Li:2019klc}
Z.~Li, H.~Gao, J.-J.~Wei, Y.-P.~Yang, B.~Zhang and Z.-H.~Zhu,
  \emph{{Cosmology-independent estimate of the fraction of baryon mass in the
  IGM from fast radio burst observations}},
  \href{https://doi.org/10.3847/1538-4357/ab18fe}{\emph{Astrophys. J.}
  {\bfseries 876} (2019) 146}
  [\href{https://arxiv.org/abs/1904.08927}{{\ttfamily 1904.08927}}].

\bibitem{Wei:2019uhh}
J.-J.~Wei, Z.~Li, H.~Gao and X.-F.~Wu, \emph{{Constraining the Evolution of the
  Baryon Fraction in the IGM with FRB and H(z) data}},
  \href{https://doi.org/10.1088/1475-7516/2019/09/039}{\emph{JCAP} {\bfseries
  09} (2019) 039} [\href{https://arxiv.org/abs/1907.09772}{{\ttfamily
  1907.09772}}].

\bibitem{Wu:2020jmx}
Q.~Wu, H.~Yu and F.Y.~Wang, \emph{{A New Method to Measure Hubble Parameter
  $H(z)$ using Fast Radio Bursts}},
  \href{https://doi.org/10.3847/1538-4357/ab88d2}{\emph{Astrophys. J.}
  {\bfseries 895} (2020) 33}
  [\href{https://arxiv.org/abs/2004.12649}{{\ttfamily 2004.12649}}].

\bibitem{Lee:2021ppm}
K.-G.~Lee, M.~Ata, I.S.~Khrykin, Y.~Huang, J.X.~Prochaska, J.~Cooke et~al.,
  \emph{{Constraining the Cosmic Baryon Distribution with Fast Radio Burst
  Foreground Mapping}},
  \href{https://doi.org/10.3847/1538-4357/ac4f62}{\emph{Astrophys. J.}
  {\bfseries 928} (2022) 9} [\href{https://arxiv.org/abs/2109.00386}{{\ttfamily
  2109.00386}}].

\bibitem{Gao:2022ifq}
R.~Gao, Z.~Li and H.~Gao, \emph{{Prospects of strongly lensed fast radio
  bursts: simultaneous measurement of post-Newtonian parameter and Hubble
  constant}}, \href{https://doi.org/10.1093/mnras/stac2270}{\emph{Mon. Not.
  Roy. Astron. Soc.} {\bfseries 516} (2022) 1977}
  [\href{https://arxiv.org/abs/2208.10175}{{\ttfamily 2208.10175}}].

\bibitem{Reischke:2023gjv}
R.~Reischke and S.~Hagstotz, \emph{{Consistent Constraints on the Equivalence
  Principle from localised Fast Radio Bursts}},
  \href{https://arxiv.org/abs/2302.10072}{{\ttfamily 2302.10072}}.

\bibitem{Bhandari:2021thi}
S.~Bhandari and C.~Flynn, \emph{{Probing the Universe with Fast Radio Bursts}},
  \href{https://doi.org/10.3390/universe7040085}{\emph{Universe} {\bfseries 7}
  (2021) 85}.

\bibitem{Xiao:2021omr}
D.~Xiao, F.~Wang and Z.~Dai, \emph{{The physics of fast radio bursts}},
  \href{https://doi.org/10.1007/s11433-020-1661-7}{\emph{Sci. China Phys. Mech.
  Astron.} {\bfseries 64} (2021) 249501}
  [\href{https://arxiv.org/abs/2101.04907}{{\ttfamily 2101.04907}}].

\bibitem{Petroff:2021wug}
E.~Petroff, J.W.T.~Hessels and D.R.~Lorimer, \emph{{Fast radio bursts at the
  dawn of the 2020s}},
  \href{https://doi.org/10.1007/s00159-022-00139-w}{\emph{Astron. Astrophys.
  Rev.} {\bfseries 30} (2022) 2}
  [\href{https://arxiv.org/abs/2107.10113}{{\ttfamily 2107.10113}}].

\bibitem{Caleb:2021xqe}
M.~Caleb and E.~Keane, \emph{{A Decade and a Half of Fast Radio Burst
  Observations}},
  \href{https://doi.org/10.3390/universe7110453}{\emph{Universe} {\bfseries 7}
  (2021) 453}.

\bibitem{CHIMEFRB:2021srp}
{\scshape CHIME/FRB} collaboration, \emph{{The First CHIME/FRB Fast Radio Burst
  Catalog}}, \href{https://doi.org/10.3847/1538-4365/ac33ab}{\emph{Astrophys.
  J. Supp.} {\bfseries 257} (2021) 59}
  [\href{https://arxiv.org/abs/2106.04352}{{\ttfamily 2106.04352}}].

\bibitem{Hagstotz:2021jzu}
S.~Hagstotz, R.~Reischke and R.~Lilow, \emph{{A new measurement of the Hubble
  constant using fast radio bursts}},
  \href{https://doi.org/10.1093/mnras/stac077}{\emph{Mon. Not. Roy. Astron.
  Soc.} {\bfseries 511} (2022) 662}
  [\href{https://arxiv.org/abs/2104.04538}{{\ttfamily 2104.04538}}].

\bibitem{Wu:2021jyk}
Q.~Wu, G.-Q.~Zhang and F.-Y.~Wang, \emph{{An 8~per\,cent determination of the
  Hubble constant from localized fast radio bursts}},
  \href{https://doi.org/10.1093/mnrasl/slac022}{\emph{Mon. Not. Roy. Astron.
  Soc.} {\bfseries 515} (2022) L1}
  [\href{https://arxiv.org/abs/2108.00581}{{\ttfamily 2108.00581}}].

\bibitem{James:2022dcx}
C.W.~James et~al., \emph{{A measurement of Hubble's Constant using Fast Radio
  Bursts}},  \href{https://arxiv.org/abs/2208.00819}{{\ttfamily 2208.00819}}.

\bibitem{Liu:2022bmn}
Y.~Liu, H.~Yu and P.~Wu, \emph{{Cosmological-model-independent determination of
  Hubble constant from fast radio bursts and Hubble parameter measurements}},
  \href{https://arxiv.org/abs/2210.05202}{{\ttfamily 2210.05202}}.

\bibitem{Zhao:2022yiv}
Z.-W.~Zhao, J.-G.~Zhang, Y.~Li, J.-M.~Zou, J.-F.~Zhang and X.~Zhang,
  \emph{{First statistical measurement of the Hubble constant using unlocalized
  fast radio bursts}},  \href{https://arxiv.org/abs/2212.13433}{{\ttfamily
  2212.13433}}.

\bibitem{Zhao:2020ole}
Z.-W.~Zhao, Z.-X.~Li, J.-Z.~Qi, H.~Gao, J.-F.~Zhang and X.~Zhang,
  \emph{{Cosmological parameter estimation for dynamical dark energy models
  with future fast radio burst observations}},
  \href{https://doi.org/10.3847/1538-4357/abb8ce}{\emph{Astrophys. J.}
  {\bfseries 903} (2020) 83}
  [\href{https://arxiv.org/abs/2006.01450}{{\ttfamily 2006.01450}}].

\bibitem{Qiu:2021cww}
X.-W.~Qiu, Z.-W.~Zhao, L.-F.~Wang, J.-F.~Zhang and X.~Zhang, \emph{{A forecast
  of using fast radio burst observations to constrain holographic dark
  energy}}, \href{https://doi.org/10.1088/1475-7516/2022/02/006}{\emph{JCAP}
  {\bfseries 02} (2022) 006}
  [\href{https://arxiv.org/abs/2108.04127}{{\ttfamily 2108.04127}}].

\bibitem{Wang:2016lxa}
B.~Wang, E.~Abdalla, F.~Atrio-Barandela and D.~Pavon, \emph{{Dark Matter and
  Dark Energy Interactions: Theoretical Challenges, Cosmological Implications
  and Observational Signatures}},
  \href{https://doi.org/10.1088/0034-4885/79/9/096901}{\emph{Rept. Prog. Phys.}
  {\bfseries 79} (2016) 096901}
  [\href{https://arxiv.org/abs/1603.08299}{{\ttfamily 1603.08299}}].

\bibitem{Comelli:2003cv}
D.~Comelli, M.~Pietroni and A.~Riotto, \emph{{Dark energy and dark matter}},
  \href{https://doi.org/10.1016/j.physletb.2003.05.006}{\emph{Phys. Lett. B}
  {\bfseries 571} (2003) 115}
  [\href{https://arxiv.org/abs/hep-ph/0302080}{{\ttfamily hep-ph/0302080}}].

\bibitem{Cai:2004dk}
R.-G.~Cai and A.~Wang, \emph{{Cosmology with interaction between phantom dark
  energy and dark matter and the coincidence problem}},
  \href{https://doi.org/10.1088/1475-7516/2005/03/002}{\emph{JCAP} {\bfseries
  03} (2005) 002} [\href{https://arxiv.org/abs/hep-th/0411025}{{\ttfamily
  hep-th/0411025}}].

\bibitem{Zhang:2005rg}
X.~Zhang, \emph{{Coupled quintessence in a power-law case and the cosmic
  coincidence problem}},
  \href{https://doi.org/10.1142/S0217732305017597}{\emph{Mod. Phys. Lett. A}
  {\bfseries 20} (2005) 2575}
  [\href{https://arxiv.org/abs/astro-ph/0503072}{{\ttfamily
  astro-ph/0503072}}].

\bibitem{He:2008tn}
J.-H.~He and B.~Wang, \emph{{Effects of the interaction between dark energy and
  dark matter on cosmological parameters}},
  \href{https://doi.org/10.1088/1475-7516/2008/06/010}{\emph{JCAP} {\bfseries
  06} (2008) 010} [\href{https://arxiv.org/abs/0801.4233}{{\ttfamily
  0801.4233}}].

\bibitem{He:2009pd}
J.-H.~He, B.~Wang and P.~Zhang, \emph{{The Imprint of the interaction between
  dark sectors in large scale cosmic microwave background anisotropies}},
  \href{https://doi.org/10.1103/PhysRevD.80.063530}{\emph{Phys. Rev. D}
  {\bfseries 80} (2009) 063530}
  [\href{https://arxiv.org/abs/0906.0677}{{\ttfamily 0906.0677}}].

\bibitem{DiValentino:2017iww}
E.~Di~Valentino, A.~Melchiorri and O.~Mena, \emph{{Can interacting dark energy
  solve the $H_0$ tension?}},
  \href{https://doi.org/10.1103/PhysRevD.96.043503}{\emph{Phys. Rev. D}
  {\bfseries 96} (2017) 043503}
  [\href{https://arxiv.org/abs/1704.08342}{{\ttfamily 1704.08342}}].

\bibitem{Yang:2018euj}
W.~Yang, S.~Pan, E.~Di~Valentino, R.C.~Nunes, S.~Vagnozzi and D.F.~Mota,
  \emph{{Tale of stable interacting dark energy, observational signatures, and
  the $H_0$ tension}},
  \href{https://doi.org/10.1088/1475-7516/2018/09/019}{\emph{JCAP} {\bfseries
  09} (2018) 019} [\href{https://arxiv.org/abs/1805.08252}{{\ttfamily
  1805.08252}}].

\bibitem{Yang:2018uae}
W.~Yang, A.~Mukherjee, E.~Di~Valentino and S.~Pan, \emph{{Interacting dark
  energy with time varying equation of state and the $H_0$ tension}},
  \href{https://doi.org/10.1103/PhysRevD.98.123527}{\emph{Phys. Rev. D}
  {\bfseries 98} (2018) 123527}
  [\href{https://arxiv.org/abs/1809.06883}{{\ttfamily 1809.06883}}].

\bibitem{Pan:2019gop}
S.~Pan, W.~Yang, E.~Di~Valentino, E.N.~Saridakis and S.~Chakraborty,
  \emph{{Interacting scenarios with dynamical dark energy: Observational
  constraints and alleviation of the $H_0$ tension}},
  \href{https://doi.org/10.1103/PhysRevD.100.103520}{\emph{Phys. Rev. D}
  {\bfseries 100} (2019) 103520}
  [\href{https://arxiv.org/abs/1907.07540}{{\ttfamily 1907.07540}}].

\bibitem{DiValentino:2019ffd}
E.~Di~Valentino, A.~Melchiorri, O.~Mena and S.~Vagnozzi, \emph{{Interacting
  dark energy in the early 2020s: A promising solution to the $H_0$ and cosmic
  shear tensions}},
  \href{https://doi.org/10.1016/j.dark.2020.100666}{\emph{Phys. Dark Univ.}
  {\bfseries 30} (2020) 100666}
  [\href{https://arxiv.org/abs/1908.04281}{{\ttfamily 1908.04281}}].

\bibitem{DiValentino:2019jae}
E.~Di~Valentino, A.~Melchiorri, O.~Mena and S.~Vagnozzi, \emph{{Nonminimal dark
  sector physics and cosmological tensions}},
  \href{https://doi.org/10.1103/PhysRevD.101.063502}{\emph{Phys. Rev. D}
  {\bfseries 101} (2020) 063502}
  [\href{https://arxiv.org/abs/1910.09853}{{\ttfamily 1910.09853}}].

\bibitem{Vagnozzi:2019ezj}
S.~Vagnozzi, \emph{{New physics in light of the $H_0$ tension: An alternative
  view}}, \href{https://doi.org/10.1103/PhysRevD.102.023518}{\emph{Phys. Rev.
  D} {\bfseries 102} (2020) 023518}
  [\href{https://arxiv.org/abs/1907.07569}{{\ttfamily 1907.07569}}].

\bibitem{Gao:2021xnk}
L.-Y.~Gao, Z.-W.~Zhao, S.-S.~Xue and X.~Zhang, \emph{{Relieving the H 0 tension
  with a new interacting dark energy model}},
  \href{https://doi.org/10.1088/1475-7516/2021/07/005}{\emph{JCAP} {\bfseries
  07} (2021) 005} [\href{https://arxiv.org/abs/2101.10714}{{\ttfamily
  2101.10714}}].

\bibitem{Zhang:2007uh}
J.~Zhang, H.~Liu and X.~Zhang, \emph{{Statefinder diagnosis for the interacting
  model of holographic dark energy}},
  \href{https://doi.org/10.1016/j.physletb.2007.10.086}{\emph{Phys. Lett. B}
  {\bfseries 659} (2008) 26} [\href{https://arxiv.org/abs/0705.4145}{{\ttfamily
  0705.4145}}].

\bibitem{Zhang:2009qa}
L.~Zhang, J.~Cui, J.~Zhang and X.~Zhang, \emph{{Interacting model of new
  agegraphic dark energy: Cosmological evolution and statefinder diagnostic}},
  \href{https://doi.org/10.1142/S0218271810016245}{\emph{Int. J. Mod. Phys. D}
  {\bfseries 19} (2010) 21} [\href{https://arxiv.org/abs/0911.2838}{{\ttfamily
  0911.2838}}].

\bibitem{Li:2009zs}
M.~Li, X.-D.~Li, S.~Wang, Y.~Wang and X.~Zhang, \emph{{Probing interaction and
  spatial curvature in the holographic dark energy model}},
  \href{https://doi.org/10.1088/1475-7516/2009/12/014}{\emph{JCAP} {\bfseries
  12} (2009) 014} [\href{https://arxiv.org/abs/0910.3855}{{\ttfamily
  0910.3855}}].

\bibitem{Li:2010ak}
Y.~Li, J.~Ma, J.~Cui, Z.~Wang and X.~Zhang, \emph{{Interacting model of new
  agegraphic dark energy: observational constraints and age problem}},
  \href{https://doi.org/10.1007/s11433-011-4382-1}{\emph{Sci. China Phys. Mech.
  Astron.} {\bfseries 54} (2011) 1367}
  [\href{https://arxiv.org/abs/1011.6122}{{\ttfamily 1011.6122}}].

\bibitem{Li:2011ga}
Y.-H.~Li and X.~Zhang, \emph{{Running coupling: Does the coupling between dark
  energy and dark matter change sign during the cosmological evolution?}},
  \href{https://doi.org/10.1140/epjc/s10052-011-1700-8}{\emph{Eur. Phys. J. C}
  {\bfseries 71} (2011) 1700}
  [\href{https://arxiv.org/abs/1103.3185}{{\ttfamily 1103.3185}}].

\bibitem{Zhang:2012uu}
Z.~Zhang, S.~Li, X.-D.~Li, X.~Zhang and M.~Li, \emph{{Revisit of the
  Interaction between Holographic Dark Energy and Dark Matter}},
  \href{https://doi.org/10.1088/1475-7516/2012/06/009}{\emph{JCAP} {\bfseries
  06} (2012) 009} [\href{https://arxiv.org/abs/1204.6135}{{\ttfamily
  1204.6135}}].

\bibitem{Zhang:2013lea}
J.~Zhang, L.~Zhao and X.~Zhang, \emph{{Revisiting the interacting model of new
  agegraphic dark energy}},
  \href{https://doi.org/10.1007/s11433-013-5378-9}{\emph{Sci. China Phys. Mech.
  Astron.} {\bfseries 57} (2014) 387}
  [\href{https://arxiv.org/abs/1306.1289}{{\ttfamily 1306.1289}}].

\bibitem{Li:2013bya}
Y.-H.~Li and X.~Zhang, \emph{{Large-scale stable interacting dark energy model:
  Cosmological perturbations and observational constraints}},
  \href{https://doi.org/10.1103/PhysRevD.89.083009}{\emph{Phys. Rev. D}
  {\bfseries 89} (2014) 083009}
  [\href{https://arxiv.org/abs/1312.6328}{{\ttfamily 1312.6328}}].

\bibitem{Li:2014eha}
Y.-H.~Li, J.-F.~Zhang and X.~Zhang, \emph{{Parametrized Post-Friedmann
  Framework for Interacting Dark Energy}},
  \href{https://doi.org/10.1103/PhysRevD.90.063005}{\emph{Phys. Rev. D}
  {\bfseries 90} (2014) 063005}
  [\href{https://arxiv.org/abs/1404.5220}{{\ttfamily 1404.5220}}].

\bibitem{Li:2014cee}
Y.-H.~Li, J.-F.~Zhang and X.~Zhang, \emph{{Exploring the full parameter space
  for an interacting dark energy model with recent observations including
  redshift-space distortions: Application of the parametrized post-Friedmann
  approach}}, \href{https://doi.org/10.1103/PhysRevD.90.123007}{\emph{Phys.
  Rev. D} {\bfseries 90} (2014) 123007}
  [\href{https://arxiv.org/abs/1409.7205}{{\ttfamily 1409.7205}}].

\bibitem{Geng:2015ara}
J.-J.~Geng, Y.-H.~Li, J.-F.~Zhang and X.~Zhang, \emph{{Redshift drift
  exploration for interacting dark energy}},
  \href{https://doi.org/10.1140/epjc/s10052-015-3581-8}{\emph{Eur. Phys. J. C}
  {\bfseries 75} (2015) 356}
  [\href{https://arxiv.org/abs/1501.03874}{{\ttfamily 1501.03874}}].

\bibitem{Li:2015vla}
Y.-H.~Li, J.-F.~Zhang and X.~Zhang, \emph{{Testing models of vacuum energy
  interacting with cold dark matter}},
  \href{https://doi.org/10.1103/PhysRevD.93.023002}{\emph{Phys. Rev. D}
  {\bfseries 93} (2016) 023002}
  [\href{https://arxiv.org/abs/1506.06349}{{\ttfamily 1506.06349}}].

\bibitem{Amendola:1999qq}
L.~Amendola, \emph{{Scaling solutions in general nonminimal coupling
  theories}}, \href{https://doi.org/10.1103/PhysRevD.60.043501}{\emph{Phys.
  Rev. D} {\bfseries 60} (1999) 043501}
  [\href{https://arxiv.org/abs/astro-ph/9904120}{{\ttfamily
  astro-ph/9904120}}].

\bibitem{Billyard:2000bh}
A.P.~Billyard and A.A.~Coley, \emph{{Interactions in scalar field cosmology}},
  \href{https://doi.org/10.1103/PhysRevD.61.083503}{\emph{Phys. Rev. D}
  {\bfseries 61} (2000) 083503}
  [\href{https://arxiv.org/abs/astro-ph/9908224}{{\ttfamily
  astro-ph/9908224}}].

\bibitem{Guo:2007zk}
Z.-K.~Guo, N.~Ohta and S.~Tsujikawa, \emph{{Probing the Coupling between Dark
  Components of the Universe}},
  \href{https://doi.org/10.1103/PhysRevD.76.023508}{\emph{Phys. Rev. D}
  {\bfseries 76} (2007) 023508}
  [\href{https://arxiv.org/abs/astro-ph/0702015}{{\ttfamily
  astro-ph/0702015}}].

\bibitem{Xia:2009zzb}
J.-Q.~Xia, \emph{{Constraint on coupled dark energy models from observations}},
  \href{https://doi.org/10.1103/PhysRevD.80.103514}{\emph{Phys. Rev. D}
  {\bfseries 80} (2009) 103514}
  [\href{https://arxiv.org/abs/0911.4820}{{\ttfamily 0911.4820}}].

\bibitem{He:2010im}
J.-H.~He, B.~Wang and E.~Abdalla, \emph{{Testing the interaction between dark
  energy and dark matter via latest observations}},
  \href{https://doi.org/10.1103/PhysRevD.83.063515}{\emph{Phys. Rev. D}
  {\bfseries 83} (2011) 063515}
  [\href{https://arxiv.org/abs/1012.3904}{{\ttfamily 1012.3904}}].

\bibitem{Fu:2011ab}
T.-F.~Fu, J.-F.~Zhang, J.-Q.~Chen and X.~Zhang, \emph{{Holographic Ricci dark
  energy: Interacting model and cosmological constraints}},
  \href{https://doi.org/10.1140/epjc/s10052-012-1932-2}{\emph{Eur. Phys. J. C}
  {\bfseries 72} (2012) 1932}
  [\href{https://arxiv.org/abs/1112.2350}{{\ttfamily 1112.2350}}].

\bibitem{Murgia:2016ccp}
R.~Murgia, S.~Gariazzo and N.~Fornengo, \emph{{Constraints on the Coupling
  between Dark Energy and Dark Matter from CMB data}},
  \href{https://doi.org/10.1088/1475-7516/2016/04/014}{\emph{JCAP} {\bfseries
  04} (2016) 014} [\href{https://arxiv.org/abs/1602.01765}{{\ttfamily
  1602.01765}}].

\bibitem{Costa:2016tpb}
A.A.~Costa, X.-D.~Xu, B.~Wang and E.~Abdalla, \emph{{Constraints on interacting
  dark energy models from Planck 2015 and redshift-space distortion data}},
  \href{https://doi.org/10.1088/1475-7516/2017/01/028}{\emph{JCAP} {\bfseries
  01} (2017) 028} [\href{https://arxiv.org/abs/1605.04138}{{\ttfamily
  1605.04138}}].

\bibitem{Feng:2016djj}
L.~Feng and X.~Zhang, \emph{{Revisit of the interacting holographic dark energy
  model after Planck 2015}},
  \href{https://doi.org/10.1088/1475-7516/2016/08/072}{\emph{JCAP} {\bfseries
  08} (2016) 072} [\href{https://arxiv.org/abs/1607.05567}{{\ttfamily
  1607.05567}}].

\bibitem{Xia:2016vnp}
D.-M.~Xia and S.~Wang, \emph{{Constraining interacting dark energy models with
  latest cosmological observations}},
  \href{https://doi.org/10.1093/mnras/stw2073}{\emph{Mon. Not. Roy. Astron.
  Soc.} {\bfseries 463} (2016) 952}
  [\href{https://arxiv.org/abs/1608.04545}{{\ttfamily 1608.04545}}].

\bibitem{Guo:2018gyo}
R.-Y.~Guo, J.-F.~Zhang and X.~Zhang, \emph{{Exploring neutrino mass and mass
  hierarchy in the scenario of vacuum energy interacting with cold dark
  matte}}, \href{https://doi.org/10.1088/1674-1137/42/9/095103}{\emph{Chin.
  Phys. C} {\bfseries 42} (2018) 095103}
  [\href{https://arxiv.org/abs/1803.06910}{{\ttfamily 1803.06910}}].

\bibitem{Li:2018ydj}
H.-L.~Li, L.~Feng, J.-F.~Zhang and X.~Zhang, \emph{{Models of vacuum energy
  interacting with cold dark matter: Constraints and comparison}},
  \href{https://doi.org/10.1007/s11433-019-9439-8}{\emph{Sci. China Phys. Mech.
  Astron.} {\bfseries 62} (2019) 120411}
  [\href{https://arxiv.org/abs/1812.00319}{{\ttfamily 1812.00319}}].

\bibitem{Feng:2019jqa}
L.~Feng, D.-Z.~He, H.-L.~Li, J.-F.~Zhang and X.~Zhang, \emph{{Constraints on
  active and sterile neutrinos in an interacting dark energy cosmology}},
  \href{https://doi.org/10.1007/s11433-019-1511-8}{\emph{Sci. China Phys. Mech.
  Astron.} {\bfseries 63} (2020) 290404}
  [\href{https://arxiv.org/abs/1910.03872}{{\ttfamily 1910.03872}}].

\bibitem{Cheng:2019bkh}
G.~Cheng, Y.-Z.~Ma, F.~Wu, J.~Zhang and X.~Chen, \emph{{Testing interacting
  dark matter and dark energy model with cosmological data}},
  \href{https://doi.org/10.1103/PhysRevD.102.043517}{\emph{Phys. Rev. D}
  {\bfseries 102} (2020) 043517}
  [\href{https://arxiv.org/abs/1911.04520}{{\ttfamily 1911.04520}}].

\bibitem{Aljaf:2020eqh}
M.~Aljaf, D.~Gregoris and M.~Khurshudyan, \emph{{Constraints on interacting
  dark energy models through cosmic chronometers and Gaussian process}},
  \href{https://doi.org/10.1140/epjc/s10052-021-09306-2}{\emph{Eur. Phys. J. C}
  {\bfseries 81} (2021) 544}
  [\href{https://arxiv.org/abs/2005.01891}{{\ttfamily 2005.01891}}].

\bibitem{Li:2020gtk}
H.-L.~Li, J.-F.~Zhang and X.~Zhang, \emph{{Constraints on neutrino mass in the
  scenario of vacuum energy interacting with cold dark matter after Planck
  2018}}, \href{https://doi.org/10.1088/1572-9494/abb7c9}{\emph{Commun. Theor.
  Phys.} {\bfseries 72} (2020) 125401}
  [\href{https://arxiv.org/abs/2005.12041}{{\ttfamily 2005.12041}}].

\bibitem{Zhang:2021yof}
M.~Zhang, B.~Wang, P.-J.~Wu, J.-Z.~Qi, Y.~Xu, J.-F.~Zhang et~al.,
  \emph{{Prospects for Constraining Interacting Dark Energy Models with 21 cm
  Intensity Mapping Experiments}},
  \href{https://doi.org/10.3847/1538-4357/ac0ef5}{\emph{Astrophys. J.}
  {\bfseries 918} (2021) 56}
  [\href{https://arxiv.org/abs/2102.03979}{{\ttfamily 2102.03979}}].

\bibitem{Lucca:2021eqy}
M.~Lucca, \emph{{Multi-interacting dark energy and its cosmological
  implications}},
  \href{https://doi.org/10.1103/PhysRevD.104.083510}{\emph{Phys. Rev. D}
  {\bfseries 104} (2021) 083510}
  [\href{https://arxiv.org/abs/2106.15196}{{\ttfamily 2106.15196}}].

\bibitem{Nunes:2022bhn}
R.C.~Nunes, S.~Vagnozzi, S.~Kumar, E.~Di~Valentino and O.~Mena, \emph{{New
  tests of dark sector interactions from the full-shape galaxy power
  spectrum}}, \href{https://doi.org/10.1103/PhysRevD.105.123506}{\emph{Phys.
  Rev. D} {\bfseries 105} (2022) 123506}
  [\href{https://arxiv.org/abs/2203.08093}{{\ttfamily 2203.08093}}].

\bibitem{Yang:2021hxg}
W.~Yang, S.~Pan, E.~Di~Valentino, O.~Mena and A.~Melchiorri, \emph{{2021-H0
  odyssey: closed, phantom and interacting dark energy cosmologies}},
  \href{https://doi.org/10.1088/1475-7516/2021/10/008}{\emph{JCAP} {\bfseries
  10} (2021) 008} [\href{https://arxiv.org/abs/2101.03129}{{\ttfamily
  2101.03129}}].

\bibitem{Guo:2021rrz}
R.-Y.~Guo, L.~Feng, T.-Y.~Yao and X.-Y.~Chen, \emph{{Exploration of interacting
  dynamical dark energy model with interaction term including the
  equation-of-state parameter: alleviation of the H$_{0}$ tension}},
  \href{https://doi.org/10.1088/1475-7516/2021/12/036}{\emph{JCAP} {\bfseries
  12} (2021) 036} [\href{https://arxiv.org/abs/2110.02536}{{\ttfamily
  2110.02536}}].

\bibitem{Wang:2021kxc}
L.-F.~Wang, J.-H.~Zhang, D.-Z.~He, J.-F.~Zhang and X.~Zhang, \emph{{Constraints
  on interacting dark energy models from time-delay cosmography with seven
  lensed quasars}}, \href{https://doi.org/10.1093/mnras/stac1468}{\emph{Mon.
  Not. Roy. Astron. Soc.} {\bfseries 514} (2022) 1433}
  [\href{https://arxiv.org/abs/2102.09331}{{\ttfamily 2102.09331}}].

\bibitem{Jin:2022tdf}
S.-J.~Jin, R.-Q.~Zhu, L.-F.~Wang, H.-L.~Li, J.-F.~Zhang and X.~Zhang,
  \emph{{Impacts of gravitational-wave standard siren observations from
  Einstein Telescope and Cosmic Explorer on weighing neutrinos in interacting
  dark energy models}},
  \href{https://doi.org/10.1088/1572-9494/ac7b76}{\emph{Commun. Theor. Phys.}
  {\bfseries 74} (2022) 105404}
  [\href{https://arxiv.org/abs/2204.04689}{{\ttfamily 2204.04689}}].

\bibitem{Majerotto:2009zz}
E.~Majerotto, J.~Valiviita and R.~Maartens, \emph{{Instability in interacting
  dark energy and dark matter fluids}},
  \href{https://doi.org/10.1016/j.nuclphysbps.2009.07.089}{\emph{Nucl. Phys. B
  Proc. Suppl.} {\bfseries 194} (2009) 260}.

\bibitem{Clemson:2011an}
T.~Clemson, K.~Koyama, G.-B.~Zhao, R.~Maartens and J.~Valiviita,
  \emph{{Interacting Dark Energy -- constraints and degeneracies}},
  \href{https://doi.org/10.1103/PhysRevD.85.043007}{\emph{Phys. Rev. D}
  {\bfseries 85} (2012) 043007}
  [\href{https://arxiv.org/abs/1109.6234}{{\ttfamily 1109.6234}}].

\bibitem{He:2008si}
J.-H.~He, B.~Wang and E.~Abdalla, \emph{{Stability of the curvature
  perturbation in dark sectors' mutual interacting models}},
  \href{https://doi.org/10.1016/j.physletb.2008.11.062}{\emph{Phys. Lett. B}
  {\bfseries 671} (2009) 139}
  [\href{https://arxiv.org/abs/0807.3471}{{\ttfamily 0807.3471}}].

\bibitem{Fang:2008sn}
W.~Fang, W.~Hu and A.~Lewis, \emph{{Crossing the Phantom Divide with
  Parameterized Post-Friedmann Dark Energy}},
  \href{https://doi.org/10.1103/PhysRevD.78.087303}{\emph{Phys. Rev. D}
  {\bfseries 78} (2008) 087303}
  [\href{https://arxiv.org/abs/0808.3125}{{\ttfamily 0808.3125}}].

\bibitem{Hu:2008zd}
W.~Hu, \emph{{Parametrized Post-Friedmann Signatures of Acceleration in the
  CMB}}, \href{https://doi.org/10.1103/PhysRevD.77.103524}{\emph{Phys. Rev. D}
  {\bfseries 77} (2008) 103524}
  [\href{https://arxiv.org/abs/0801.2433}{{\ttfamily 0801.2433}}].

\bibitem{Zhang:2017ize}
X.~Zhang, \emph{{Probing the interaction between dark energy and dark matter
  with the parametrized post-Friedmann approach}},
  \href{https://doi.org/10.1007/s11433-017-9013-7}{\emph{Sci. China Phys. Mech.
  Astron.} {\bfseries 60} (2017) 050431}
  [\href{https://arxiv.org/abs/1702.04564}{{\ttfamily 1702.04564}}].

\bibitem{Feng:2018yew}
L.~Feng, Y.-H.~Li, F.~Yu, J.-F.~Zhang and X.~Zhang, \emph{{Exploring
  interacting holographic dark energy in a perturbed universe with
  parameterized post-Friedmann approach}},
  \href{https://doi.org/10.1140/epjc/s10052-018-6338-3}{\emph{Eur. Phys. J. C}
  {\bfseries 78} (2018) 865}
  [\href{https://arxiv.org/abs/1807.03022}{{\ttfamily 1807.03022}}].

\bibitem{Shull:2011aa}
J.M.~Shull, B.D.~Smith and C.W.~Danforth, \emph{{The Baryon Census in a
  Multiphase Intergalactic Medium: 30\% of the Baryons May Still Be Missing}},
  \href{https://doi.org/10.1088/0004-637X/759/1/23}{\emph{Astrophys. J.}
  {\bfseries 759} (2012) 23} [\href{https://arxiv.org/abs/1112.2706}{{\ttfamily
  1112.2706}}].

\bibitem{Fan:2006dp}
X.-H.~Fan, C.L.~Carilli and B.G.~Keating, \emph{{Observational constraints on
  cosmic reionization}},
  \href{https://doi.org/10.1146/annurev.astro.44.051905.092514}{\emph{Ann. Rev.
  Astron. Astrophys.} {\bfseries 44} (2006) 415}
  [\href{https://arxiv.org/abs/astro-ph/0602375}{{\ttfamily
  astro-ph/0602375}}].

\bibitem{McQuinn:2013tmc}
M.~McQuinn, \emph{{Locating the ''missing'' baryons with extragalactic
  dispersion measure estimates}},
  \href{https://doi.org/10.1088/2041-8205/780/2/L33}{\emph{Astrophys. J. Lett.}
  {\bfseries 780} (2014) L33}
  [\href{https://arxiv.org/abs/1309.4451}{{\ttfamily 1309.4451}}].

\bibitem{Reischke:2023blu}
R.~Reischke and S.~Hagstotz, \emph{{Covariance Matrix of Fast Radio Bursts
  Dispersion}},  \href{https://arxiv.org/abs/2301.03527}{{\ttfamily
  2301.03527}}.

\bibitem{Zhang:2020mgq}
G.Q.~Zhang, H.~Yu, J.H.~He and F.Y.~Wang, \emph{{Dispersion measures of fast
  radio burst host galaxies derived from IllustrisTNG simulation}},
  \href{https://doi.org/10.3847/1538-4357/abaa4a}{\emph{Astrophys. J.}
  {\bfseries 900} (2020) 170}
  [\href{https://arxiv.org/abs/2007.13935}{{\ttfamily 2007.13935}}].

\bibitem{Hashimoto:2020dud}
T.~Hashimoto, T.~Goto, A.Y.L.~On, T.-Y.~Lu, D.J.D.~Santos, S.C.C.~Ho et~al.,
  \emph{{Fast radio bursts to be detected with the Square Kilometre Array}},
  \href{https://doi.org/10.1093/mnras/staa2238}{\emph{Mon. Not. Roy. Astron.
  Soc.} {\bfseries 497} (2020) 4107}
  [\href{https://arxiv.org/abs/2008.00007}{{\ttfamily 2008.00007}}].

\bibitem{ZJG}
J.-G.~Zhang, Z.-W.~Zhao, Y.-C.~Li, J.-F.~Zhang, D.~Li and X.~Zhang,
  \emph{{Cosmology with fast radio bursts in the era of SKA}}, {\emph{in
  preparation} }.

\bibitem{Zhang:2021kdu}
R.C.~Zhang and B.~Zhang, \emph{{The CHIME Fast Radio Burst Population Does Not
  Track the Star Formation History of the Universe}},
  \href{https://doi.org/10.3847/2041-8213/ac46ad}{\emph{Astrophys. J. Lett.}
  {\bfseries 924} (2022) L14}
  [\href{https://arxiv.org/abs/2109.07558}{{\ttfamily 2109.07558}}].

\bibitem{Qiang:2021ljr}
D.-C.~Qiang, S.-L.~Li and H.~Wei, \emph{{Fast radio burst distributions
  consistent with the first CHIME/FRB catalog}},
  \href{https://doi.org/10.1088/1475-7516/2022/01/040}{\emph{JCAP} {\bfseries
  01} (2022) 040} [\href{https://arxiv.org/abs/2111.07476}{{\ttfamily
  2111.07476}}].

\bibitem{Madau:2016jbv}
P.~Madau and T.~Fragos, \emph{{Radiation Backgrounds at Cosmic Dawn: X-Rays
  from Compact Binaries}},
  \href{https://doi.org/10.3847/1538-4357/aa6af9}{\emph{Astrophys. J.}
  {\bfseries 840} (2017) 39}
  [\href{https://arxiv.org/abs/1606.07887}{{\ttfamily 1606.07887}}].

\bibitem{Chen:2018dbv}
L.~Chen, Q.-G.~Huang and K.~Wang, \emph{{Distance Priors from Planck Final
  Release}}, \href{https://doi.org/10.1088/1475-7516/2019/02/028}{\emph{JCAP}
  {\bfseries 02} (2019) 028}
  [\href{https://arxiv.org/abs/1808.05724}{{\ttfamily 1808.05724}}].

\bibitem{Beutler:2011hx}
F.~Beutler, C.~Blake, M.~Colless, D.H.~Jones, L.~Staveley-Smith, L.~Campbell
  et~al., \emph{{The 6dF Galaxy Survey: Baryon Acoustic Oscillations and the
  Local Hubble Constant}},
  \href{https://doi.org/10.1111/j.1365-2966.2011.19250.x}{\emph{Mon. Not. Roy.
  Astron. Soc.} {\bfseries 416} (2011) 3017}
  [\href{https://arxiv.org/abs/1106.3366}{{\ttfamily 1106.3366}}].

\bibitem{Ross:2014qpa}
A.J.~Ross, L.~Samushia, C.~Howlett, W.J.~Percival, A.~Burden and M.~Manera,
  \emph{{The clustering of the SDSS DR7 main Galaxy sample \textendash{} I. A 4
  per cent distance measure at $z = 0.15$}},
  \href{https://doi.org/10.1093/mnras/stv154}{\emph{Mon. Not. Roy. Astron.
  Soc.} {\bfseries 449} (2015) 835}
  [\href{https://arxiv.org/abs/1409.3242}{{\ttfamily 1409.3242}}].

\bibitem{Alam:2016hwk}
{\scshape BOSS} collaboration, \emph{{The clustering of galaxies in the
  completed SDSS-III Baryon Oscillation Spectroscopic Survey: cosmological
  analysis of the DR12 galaxy sample}},
  \href{https://doi.org/10.1093/mnras/stx721}{\emph{Mon. Not. Roy. Astron.
  Soc.} {\bfseries 470} (2017) 2617}
  [\href{https://arxiv.org/abs/1607.03155}{{\ttfamily 1607.03155}}].

\bibitem{Pan-STARRS1:2017jku}
{\scshape Pan-STARRS1} collaboration, \emph{{The Complete Light-curve Sample of
  Spectroscopically Confirmed SNe Ia from Pan-STARRS1 and Cosmological
  Constraints from the Combined Pantheon Sample}},
  \href{https://doi.org/10.3847/1538-4357/aab9bb}{\emph{Astrophys. J.}
  {\bfseries 859} (2018) 101}
  [\href{https://arxiv.org/abs/1710.00845}{{\ttfamily 1710.00845}}].

\bibitem{Lewis:1999bs}
A.~Lewis, A.~Challinor and A.~Lasenby, \emph{{Efficient computation of CMB
  anisotropies in closed FRW models}},
  \href{https://doi.org/10.1086/309179}{\emph{Astrophys. J.} {\bfseries 538}
  (2000) 473} [\href{https://arxiv.org/abs/astro-ph/9911177}{{\ttfamily
  astro-ph/9911177}}].

\bibitem{Lewis:2002ah}
A.~Lewis and S.~Bridle, \emph{{Cosmological parameters from CMB and other data:
  A Monte Carlo approach}},
  \href{https://doi.org/10.1103/PhysRevD.66.103511}{\emph{Phys. Rev. D}
  {\bfseries 66} (2002) 103511}
  [\href{https://arxiv.org/abs/astro-ph/0205436}{{\ttfamily
  astro-ph/0205436}}].

\bibitem{KochOcker:2022ook}
S.~Koch~Ocker et~al., \emph{{The Large Dispersion and Scattering of FRB
  20190520B Are Dominated by the Host Galaxy}},
  \href{https://doi.org/10.3847/1538-4357/ac6504}{\emph{Astrophys. J.}
  {\bfseries 931} (2022) 87}
  [\href{https://arxiv.org/abs/2202.13458}{{\ttfamily 2202.13458}}].

\bibitem{Masui:2015ola}
K.W.~Masui and K.~Sigurdson, \emph{{Dispersion Distance and the Matter
  Distribution of the Universe in Dispersion Space}},
  \href{https://doi.org/10.1103/PhysRevLett.115.121301}{\emph{Phys. Rev. Lett.}
  {\bfseries 115} (2015) 121301}
  [\href{https://arxiv.org/abs/1506.01704}{{\ttfamily 1506.01704}}].

\bibitem{Reischke:2020cgd}
R.~Reischke, S.~Hagstotz and R.~Lilow, \emph{{Probing primordial
  non-Gaussianity with Fast Radio Bursts}},
  \href{https://doi.org/10.1103/PhysRevD.103.023517}{\emph{Phys. Rev. D}
  {\bfseries 103} (2021) 023517}
  [\href{https://arxiv.org/abs/2007.04054}{{\ttfamily 2007.04054}}].

\bibitem{Shirasaki:2021rzq}
M.~Shirasaki, R.~Takahashi, K.~Osato and K.~Ioka, \emph{{Probing cosmology and
  gastrophysics with fast radio bursts:~cross-correlations of dark matter
  haloes and cosmic dispersion measures}},
  \href{https://doi.org/10.1093/mnras/stac490}{\emph{Mon. Not. Roy. Astron.
  Soc.} {\bfseries 512} (2022) 1730}
  [\href{https://arxiv.org/abs/2108.12205}{{\ttfamily 2108.12205}}].

\bibitem{Rafiei-Ravandi:2021hbw}
M.~Rafiei-Ravandi et~al., \emph{{CHIME/FRB Catalog 1 Results: Statistical
  Cross-correlations with Large-scale Structure}},
  \href{https://doi.org/10.3847/1538-4357/ac1dab}{\emph{Astrophys. J.}
  {\bfseries 922} (2021) 42}
  [\href{https://arxiv.org/abs/2106.04354}{{\ttfamily 2106.04354}}].

\end{thebibliography}\endgroup
\bibliographystyle{JHEP}

\end{document}